\documentclass[10pt,oneside]{article}

\usepackage[small,bf,margin=0.5cm]{caption}
\usepackage{subfig}

\usepackage{graphicx,color}
\usepackage{epstopdf}
\usepackage{pgf}
\usepackage{pgffor}

\usepackage{amsmath}
\usepackage{latexsym}
\usepackage{amssymb}
\usepackage{amsfonts}
\usepackage{pifont}
\usepackage{tikz}
\usetikzlibrary{shapes,arrows}
\usetikzlibrary{calc}
\usepackage[english]{babel}
\usepackage{hyperref}
\usepackage[all]{hypcap}
\usepackage{lastpage}
\usepackage{fullpage}
\usepackage{blkarray}
\usepackage{amsthm}

\usepackage[affil-it]{authblk}
\usepackage[top=.75in,left=.9in,bottom=1.in,right=.9in]{geometry}
\usepackage{tensor}
\usepackage{bm}
\usepackage{comment}

    \makeatletter
     
      \@addtoreset{equation}{subsection}
    \makeatother

     \makeatletter
        
        \@addtoreset{figure}{section}
     \makeatother

\addto\captionsenglish{}

\begin{document}

\title{On-the-fly algorithm for Dynamic Mode Decomposition using Incremental Singular Value Decomposition and Total Least Squares}
%
\author[1]{Daiki~Matsumoto\thanks{Corresponding author: \href{mailto:daiki.matsumoto@aer.mw.tum.de}{daiki.matsumoto@aer.mw.tum.de}}}
\author[1]{Thomas~Indinger}
\affil[1]{{{Chair of Aerodynamics and Fluid Mechanics, Technical University of Munich, \protect\\ Boltzmann.str 15, 85748, Garching b. Muenchen, Germany}}}
\date{}

\maketitle

\begin{abstract}
Dynamic Mode Decomposition (DMD) is a useful tool to effectively extract the dominant dynamic flow structure from a unsteady flow field. However, DMD requires massive computational resources with respect to memory consumption and the usage of storage.
In this paper, an alternative incremental algorithm of Total DMD (Incremental TDMD) is proposed which is based on Incremental Singular Value Decomposition (SVD).
The advantage of Incremental TDMD compared to the existing on-the-fly algorithms of DMD is that Sparsity-Promoting DMD (SPDMD) can be performed after the incremental process without saving huge datasets on the disk space.
SPDMD combined with Incremental TDMD enable the effective identification of dominant modes which are relevant to the results from conventional TDMD combined with SPDMD.
\end{abstract}

\section{Introduction}

With the development of the performance of computers, unsteady Computational Fluid Dynamics (CFD) is getting to be widely used for industrial applications such as the analysis of unsteady aerodynamics of road vehicles. 
However, it is usually challenging to identify the dominant characteristic unsteady flow structure around complex shaped objects such as a road vehicle, as the flow field consists of various flow structures with respect to time and length scale.
Dynamic Mode Decomposition (DMD)~\cite{Rowley2009,Schmid2010} seems to be very useful tool to extract dominant dynamic flow structures from the complex flow field and its usage increases significantly in recent years.

However, there are still some issues in the conventional DMD algorithm proposed by Schmid~\cite{Schmid2010}, especially for industrial applications.
One issue is that the conventional DMD algorithm is very sensitive to the noise existing in the flow field~\cite{Bagheri2014,Hemati2015,Dawson2016}.
For this issue, Hemati, et al.~\cite{Hemati2015} proposes a noise-robust DMD algorithm called Total DMD (Total-least squared DMD) which solves a total least-square problem instead of a least-square problem implicitly included in the conventional DMD algorithm.
It is reported that this Total DMD algorithm effectively removes the noise and result in better DMD modes compared to the conventional algorithms.~\cite{Hemati2015,Dawson2016}

Another issue is that many DMD modes are obtained by DMD computation, when the target flow field is very complex.  Then, it is very difficult to identify the dominant DMD modes from all the obtained modes. 
Many publications indicate that the importance of a DMD mode is judged by the amplitude of the DMD mode computed, according to ~\cite{Rowley2009} by assuming the DMD modes are good approximation of Koopman modes.
However, it is also reported that damped DMD modes often give very large amplitude, which affect the proper identification of dominant modes~\cite{Jovanovic2014,Tu2014}.
Therefore, Jovanovi\'c, et al., ~\cite{Jovanovic2014} propose the optimal amplitude of DMD which is computed by solving the convex optimization problem of the minimizing error norm between the original dataset and the reconstructed dataset from DMD modes.
In addition, a penalty parameter is introduced to the convex optimization problem in ~\cite{Jovanovic2014}, which selects some important modes from all the obtained modes with respect to the contribution for the approximation of the original dataset by remained DMD modes.
It is the trade-off method between the accuracy of the approximation by DMD modes and the number of remained DMD modes, but it seems to be very useful to identify the dominant modes from many computed modes, especially for the case of a very complex flow field and in industrial applications. 

In engineering applications of DMD, one issue is that most DMD algorithms require massive computational resources with respect to memory consumption and data storage, since all datasets of the flow field used for the DMD process have to be saved during the CFD simulation and loaded on the memory space for computing the modes.
However, it seems that using the current DMD process for an actual industrial application, especially for the development of new products, is unrealistic. 
It is because the results of CFD on engineering applications are represented by the high dimensional state of space, where, for example, the numbers of computational cells have already reached $10^{8} \sim 10^{11}$ cells for the case of an unsteady CFD of a road vehicle. 
On-the-fly DMD algorithms such as Streaming DMD (SDMD)~\cite{Hemati2014} and Streaming Total DMD (STDMD)~\cite{Hemati2016} seem to be very useful for engineering applications, as SDMD and STDMD can reduce memory consumption and it is not necessary to save flow field datasets on the storage.
These on-the-fly algorithms of DMD compute matrices incrementally when the new snapshot vector is obtained from CFD, therefore, it is not necessary to save datasets of the flow field on the storage, and can be performed in parallel to the CFD simulation.
In addition, matrices used for on-the-fly DMD algorithms are compressed after updated incrementally, which results in much less memory consumption than conventional DMD.
However, there is one issue in algorithms of SDMD and STDMD with respect to the identification of dominant DMD modes from obtained modes.
If many modes are required for the relevant computation of DMD, it seems to be still difficult to identify the dominant modes after DMD computation.
However, SPDMD introduced above cannot be used after SDMD and STDMD computations, because of the lack of information for constructing the convex optimization problem.

Therefore, in this paper, we propose an alternative on-the-fly algorithm of DMD, which can be performed with SPDMD.
To achieve this algorithm, first, the alternative algorithm of Total DMD (Alternative TDMD) is proposed, which can be computed only by singular vectors and values of augmented snapshot matrix of the flow field.
Then, Alternative TDMD can be performed on the fly, when Incremental Singular Value Decomposition (Incremental SVD)~\cite{Brand2002, Brand2006, Oxberry2017} is performed instead of the conventional SVD (Incremental TDMD).
The advantage of the Incremental TDMD algorithm compared to SDMD and STDMD is that SPDMD can be performed after the incremental process, which seems to be useful for the identification of dominant modes, even if many modes are obtained by the on-the-fly DMD computation.
Therefore, Incremental TDMD can be performed in parallel to CFD with less memory, without saving snapshots on the disk, which result in an efficient computation of DMD and easier identification of dominant modes in the flow field by applying SPDMD after incremental process.

In this paper, the flow field around the infinite square cylinder is simulated as a test case of DMD computations for validation purpose.
First, Alternative TDMD with SPDMD is performed on the flow field and compared with conventional TDMD with SPDMD, in order to validate the Alternative TDMD algorithm.
Second, the Incremental TDMD with SPDMD is applied on the same flow field and the relevancy of the Incremental TDMD algorithm for the identification of dominant modes compared to the conventional TDMD algorithms is investigated.
%
%
%


\section{Conventional Dynamic Mode Decomposition Methods}
In this section, algorithms of so-called conventional Dynamic Mode Decomposition (DMD)~\cite{Rowley2009, Schmid2010} are introduced, following descriptions of Schmid~\cite{Schmid2010} and Tu, et al.~\cite{TuThesis}.
Additionally, conventional Total DMD  (TDMD, \emph{total least-square} DMD, or \emph{tls} DMD)~\cite{Hemati2015,Dawson2016}, which is more noise-robust algorithm of DMD, is also introduced. 

Next, the scaling method of Dynamic Modes, which is useful to investigate the contribution of each DMD mode to the original flow field before decomposed, is introduced. One of the scaling method is introduced by Rowley, et al., where scaling factor (namely DMD mode amplitude) is computed by using the first snapshot of velocity field.
This scaling method is so far widely used, however, some issues are also reported in~\cite{Jovanovic2014, Tu2014}. Therefore, Sparsity-promoting DMD (SPDMD) is proposed by Jovanovi\'c, et al.~\cite{Jovanovic2014}.
SPDMD compute optimal amplitudes by solving convex optimization problem of minimizing error norm.
In addition, Jovanovi\'c et al.,~\cite{Jovanovic2014} introduced penalty parameter which induces sparsity structure of amplitude vector, and chose relatively important modes from all of obtained mode, which seem to be very useful to identify the dominant modes and flow structures from complex flow field.

Please note that, many SVD operations are included in DMD algorithms. However, the method of snapshot~\cite{Sirovich1987}, instead of conventional SVD, is adopted in this paper, to compute singular values and vectors in order to save memory and computational time for DMD computations, if there is no specification.
For example, the conventional DMD algorithm performed by the method of snapshot is explained in the Appendix.A.
%
%

\subsection{Conventional Dynamic Mode Decomposition}
Dynamic Mode Decomposition (DMD) is a data-driven method of modal analysis which can extract dominant, coherent flow structures from complex, unsteady flow field.
The objective of DMD is extracting dynamic information and its spatial structure described by eigenvectors and eigenvalues of DMD operator~${\bm A}$ which is described as follows, 
%
\begin{equation}
  {\bm Y} = {\bm A} {\bm X}
  \label{eq:LinearEq}
\end{equation}
where $ {\bm X }=\{ {\bm x_{0}}, {\bm x_{1}},\ldots,{\bm x_{m-1}}\} \in \mathbb{R}^{n \times m}$ and ${\bm Y} =\{ {\bm x_{1}}, {\bm x_{2}},\ldots,{\bm x_{m}}\} \in \mathbb{R}^{n \times m}$ are snapshot matrices consisting of snapshots of the flow field~${\bm x_{i}} \in \mathbb{R}^{n}$ which is recorded at a constant time interval ${\Delta}t_{DMD}$.
One solution to compute the DMD operator~${\bm A} \in \mathbb{R}^{n \times n}$ is that solving the following equation. ($ {\bm X}^{+}$ represents Moor-Penrose pseudo-inverse of~${\bm X}$)
%
\begin{equation}
  {\bm A} = {\bm Y} {\bm X}^{+}
  \label{eq:DMDOperator}
\end{equation}
However, it is usually difficult to derive DMD operator A and computing the eigendecomposition of~${\bm A}$, as the number of states described by~$n$ is much bigger than the number of snapshots~$m$ in many cases of unsteady CFD in the field of applied aerodynamics.
Therefore, the most of DMD algorithm attempt to approximate eigenvectors and eigenvalues of~${\bm A}$ efficiently, with small computational resources. 

Schmid, et al.~\cite{Schmid2010}, proposed the efficient algorithm to compute the eigendecomposition of A in low-dimensional approximation by using Singular Value Decomposition (SVD), which is so far the most widely used as DMD algorithm.
In this algorithm, DMD operator~${\bm A }$ is projected on the left singular vectors of~${\bm X}$ which is expressed as~$\bm U$, then projected DMD operator $\tilde{{\bm A}}$ is defined as follows, 
%
\begin{equation}
  \tilde{\bm A} = {\bm U}^{T} {\bm A} {\bm U}
  \label{eq:DefPrjDMDOperator}
\end{equation}
where the left singular vectors~$\bm U$ are computed by SVD on~$\bm X$, as follows.
%
\begin{equation}
  {\bm X} = {\bm U} {\bm \Sigma} {\bm W}^{T}
  \label{eq:SVD_X}
\end{equation}
Please note that the left singular vectors~$\bm U$ represent the modes of Proper Orthogonal Decomposition (POD)~\cite{Lumley1967} of~$\bm X$. Therefore, truncating the higher ranks of POD modes when projecting (Eq.\eqref{eq:DefPrjDMDOperator}), can work as a kind of filtering. It is because it is considered that the higher ranks of POD modes have less contribution to the original flow. Substituting Eq.\eqref{eq:SVD_X} in Eq.\eqref{eq:DefPrjDMDOperator}, the projected DMD operator can be rewritten as follows.
%
%
%
\begin{equation}
  \tilde{{\bm A}} = {\bm U}^{T} {\bm Y} {\bm X}^{+} {\bm U} =  {\bm U}^{T} {\bm Y} {\bm W} {{\bm \Sigma}}^{-1} 
  \label{eq:PrjDMDOperator_Schmid}
\end{equation}
where pseudo-inverse of~$\bm X$ is computed by using matrices derived by SVD.
%
\begin{equation}
  {\bm X}^{+} = {\bm W} {{\bm \Sigma}}^{-1} {\bm U}^{T}
  \label{eq:pinv_X}
\end{equation}
Then, eigendecomposition of projected DMD operator~$\tilde{\bm A}$ is performed as follows.
%
\begin{equation}
  \tilde{{\bm A}} {{\bm V}} = {{\bm V}} {\bm \Lambda}
  \label{eq:eig_prjAtilde}
\end{equation}
Finally, unscaled DMD modes~$\hat{{\bm \Phi}} = \{{\hat{{\bm \phi}}_{0}}, {\hat{{\bm \phi}}_{1}}, \ldots,  {\hat{{\bm \phi}}_{r-1}}\}$ are computed by projecting the eigenvectors ~${\bm V} = \{{\bm {v}_{0}}, {\bm {v}_{1}}, \ldots, {\bm {v}_{r-1}}\}$ on the left singular vectors~$\bm U$.
%
\begin{equation}
  \hat{{\bm \Phi}} = {\bm {U V}}
  \label{eq:unscaledDMDmode}
\end{equation}
DMD modes describe spatial structure of the flow field extracted by DMD.
Eigenvalues~${\bm \Lambda} = diag\{\lambda_{0}, \lambda_{1},\ldots, \lambda_{r-1}\}$ describe the amplification and damping rate and frequency of the flow structures extracted by the corresponding DMD modes. Especially, frequency of the k-th DMD mode is computed as follows.
%
\begin{equation}
  f_k = \frac{Im({\log {{\lambda}_k}})}{2 \pi {{\Delta}t}_{DMD}}
  \label{eq:FrequencyDMD}
\end{equation}

Again, DMD modes and their eigenvalues, which are computed through DMD operator, are regarded as low-dimensional approximation of eigenvectors and eigenvalues of DMD operator~$\bm A$.
According to the description by Tu., et al.~\cite{TuThesis}, it can be explained by the following process.
Firstly, substituting Eq.\eqref{eq:DefPrjDMDOperator} into Eq.\eqref{eq:eig_prjAtilde}, eigendecomposition of projected DMD operator can be expressed as follows.
%
\begin{equation}
  ({\bm U}^{T} {\bm A} {\bm U}) {\bm V} = {\bm V} {\bm \Lambda}
  \label{eq:Eq2110}
\end{equation}
Please note that left singular vectors $\bm U$ is orthogonal basis, therefore multiplying~${\bm U}^{T} {\bm U} = {\bm I}$ to the right side of Eq.\eqref{eq:Eq2110} doesn't change the balance between left and right in the equation, then
%
\begin{equation}
  {\bm U}^{T} {\bm A} {\bm U} {\bm V} = {\bm U}^{T} {\bm U} {\bm V} {\bm \Lambda}
  \label{eq:Eq2111}
\end{equation}
If pseudo-inverse of ${\bm U}^{T}$ can be multiplied from left direction for left and right side of Eq.\eqref{eq:Eq2111}, then eigendecomposition of DMD operator $\bm A$ can be derived, by recognizing $\hat{{\bm \Phi}} = {\bm {U V}}$ are eigenvectors of ${\bm A}$, as follows.
%
\begin{equation}
\begin{split}
  {\bm A} \hat{{\bm \Phi}} = \hat{{\bm \Phi}} {\bm \Lambda} \\
  where~\hat{{\bm \Phi}} = {\bm U} {\bm V}
  \label{eq:Eq2112}
\end{split}
\end{equation}
Hence, in other words, it is believed that any orthogonal basis can be used for constructing projecting DMD operator and for DMD computation, as long as the dimension is consistent in the algorithm.

\subsection{Conventional Total DMD}
In standard DMD algorithm, least-square problem is implicitly included, namely Eq.\eqref{eq:DMDOperator} is regarded as the \emph{least-squares} solution to minimize as shown in the following.
%
\begin{equation}
  \min_{{\bm A}, \Delta {\bm Y}} \|\Delta {\bm Y}\|_F,~~subject~to ~~ {\bm Y} + \Delta {\bm Y} = {\bm {AX}},
    \label{eq:LeastSquareDMD}
\end{equation}
where noise existing in snapshot matrix~$\bm Y$ is represented as~${\Delta} \bm Y$.  Therefore, DMD finds a linear relationship between the snapshots~$\bm X$ and the so-called noise-free snapshots~$\bm Y$ where noise~${\Delta} \bm Y$ is removed. Therefore, noise existing in~$\bm X$ is not considered in conventional DMD algorithm, which means~$\bm X$ and~$\bm Y$ are treated asymmetrically. This asymmetry induces a bias in the eigenvalues of~$\bm A$.

In order to account for the noise in snapshot matrix~$\bm X$ as well as~$\bm Y$,  Hemati, et al.,~\cite{Hemati2015} proposed to solve the following “\emph{total-least-square} problem”.
%
\begin{equation}
  \min_{{\bm A}, \Delta {\bm X}, \Delta {\bm Y}}\left\|
    \begin{bmatrix}
      \Delta {\bm X}\\
      \Delta {\bm Y}
    \end{bmatrix}
  \right\|_F, ~~subject~to~~{\bm Y} + \Delta {\bm Y} = {\bm A}({\bm X} + \Delta {\bm X}).
    \label{eq:TotalLeastSquareDMD}  
\end{equation}
In Total DMD algorithm, above \emph{total-least-squares} problem is solved by the projection of snapshot matrices on the best ‘r-dimensional’ subspace determined by truncated SVD on augmented snapshot matrix~$\bm Z$ which is represented as follows.  
%
\begin{equation}
  \bm Z = \left[
        \begin{array}{c}
          \bm X \\
          \bm Y
        \end{array}
      \right]
    = {\bm{U_{Z}}}_{:,1:r} {\bm{{\Sigma}_{Z}}}_{1:r} {{\bm{W_{Z}}}_{:,1:r}}^{T} 
  \label{eq:SVD_Z_TLS}
\end{equation}
Snapshot matrices~$\bm X$ and~$\bm Y$ are projected on the subspace determined by the augmented matrix~$\bm Z$, which result in namely noise-free snapshot dataset~$\overline{\bm X}$ and~$\overline{\bm Y}$ as follows.
%
\begin{equation}
\begin{split}
      \overline{\bm X} = \bm X \bm{W_{Z}} {\bm{W_{Z}}}^{T} \\
      \overline{\bm Y} = \bm Y \bm{W_{Z}} {\bm{W_{Z}}}^{T}
        \label{eq:TLSProj}
\end{split}
\end{equation}
This projection step can work for de-biasing any DMD-like algorithms. Then, DMD algorithm can be performed on ‘noise-free’ snapshot matrices computed by Eq.\eqref{eq:TLSProj}.
%
%


\subsection{Scaling of DMD modes and Reconstruction}
 Considering that DMD modes are the approximation of Koopman modes according to \cite{Rowley2009}, the i-th flow field~$\bm x_{i}$ can be approximated by scaled DMD modes~${\bm \phi}_{k}$ and corresponding eigenvalues~$\lambda_{k}$, as follows.
%
\begin{equation}
    {\bm x_{i}} = \sum_{k=0}^{r-1}{{\lambda_{k}}^{i} {\bm \phi}_{k}},~~i=0,\ldots,m-1
    \label{eq:ReconsDMD}
\end{equation}
where k-th DMD mode is scaled by the scaling factor (namely, DMD mode amplitude) $d_{k}$
%
\begin{equation}
    {\bm \phi}_{k} = d_k \hat{{\bm \phi}_{k}}
    \label{eq:ScaleDMDmode}
\end{equation}
In the different points of view, snapshot matrix of original flow field is also approximated as the following in matrix form.
%
\begin{equation}
    {\bm X} = \hat{\bm \Phi} \bm{D_{\alpha}} {\bm T} = \bm {U V D_{\alpha} T}
    \label{eq:ReconsFlowFromDMD}
\end{equation}
where 
%
\begin{equation}
    \bm T = \left[
        \begin{array}{ccc}
          {{{\lambda}_0}^{0}} & \ldots & {{{\lambda}_0}^{m-2}} \\
          \vdots & \ddots  & \vdots \\
          {{{\lambda}_{r-1}}^{0}} & \ldots & {{{\lambda}_{r-1}}^{m-2}}
        \end{array}
    \right]
    \label{eq:VandermodeMatrix}
\end{equation}
Matrix~$\bm{T}$ is called Vandermonde matrix governing the temporal evolution of dynamic modes.\cite{Rowley2009}

The solution to identify the scaling factor of DMD modes is that assuming the first snapshot of the flow field~${\bm x}_{0}$ can be expressed as a linear combination of scaled DMD modes with an assumption that DMD modes are good approximation of Koopman modes.
\begin{equation}
    {\bm x}_{0} = \sum_{k=0}^{r-1} {{\bm \phi}_{k}}
    \label{eq:ReconsDMD_1stSnapshot}
\end{equation}
Scaling factor ${\bm d} = \{{d}_{0}, {d}_{1}, \ldots, {d}_{k}\}$ (where $\bm{D_{\alpha}} = diag(\bm d)$) is derived by solving the following linear equation.
\begin{equation}
    \hat{{\bm \Phi}}{\bm d} = {\bm x}_{0}
    \label{eq:LinearEq_Scaling}
\end{equation}
One solution of Eq.\eqref{eq:LinearEq_Scaling} is that the pseudo-inverse of the unscaled DMD modes is computed and multiplied to the first snapshot vector, as follows.
%
\begin{equation}
    {\bm d} = {\hat{{\bm \Phi}}}^{+} {\bm x}_{0}
    \label{eq:Solution_LinearEq_Scaling}
\end{equation}
When the scaling factors~$\bm d$ is computed by Eq.\eqref{eq:Solution_LinearEq_Scaling}, the i-th flow field of~${\bm x}_{i}$ is reconstructed from scaled DMD modes (in Eq.\eqref{eq:ScaleDMDmode}) and corresponding eigenvalue as shown in Eq.\eqref{eq:ReconsDMD}.
Additionally, the scaling factors are often used for the identification of the dominant DMD modes, as expressing the amplitudes of the fluctuation of mode.

\subsection{Sparsity-Promoting DMD}
It is reported that amplitudes of very damped modes are computed high and it is sometimes very difficult to identify the dominant mode by the DMD mode amplitude computed from first snapshot vector as described in the section 2.4~\cite{Jovanovic2014,Tu2014}.
In addition, if many modes are obtained by DMD computation, it is more difficult to identify dominant modes representing dominant flow structures, which can easily happen for the case of very complex flow field in engineering applications.
Sparsity-promoting DMD (SPDMD) algorithm proposed by Jovanovi\'c et al.,~\cite{Jovanovic2014} is the method to achieve a desirable trade-off between the quality of approximation in the sense of least-squares and the number of modes, by using convex optimization method.
In this algorithm, the norm of entire flow field is considered to compute optimal amplitude, therefore, the effect of damping on the result of amplitude seems be more effectively removed.

With respect to the approximation of the original flow field by the reconstruction from DMD modes, the optimal amplitude of DMD modes are approximated by minimizing the error norm between snapshot matrix of the original flow field and the reconstructed flow field from DMD modes, as follows.
%
\begin{equation}
  \min_{\bm \alpha} {\| \bm X - {\hat{\bm \Phi}} \bm {D_{\alpha}} \bm T\|_F} ^{2}
    \label{eq:MinErrorNorm_ReconsFlow}
\end{equation}
The original flow field~$\bm X$ is decomposed by SVD such as Eq.\eqref{eq:SVD_X} and the reconstructed flow field from DMD is represented as Eq.\eqref{eq:ReconsFlowFromDMD} with optimal amplitudes $\bm{D_{\alpha}} = diag({\bm \alpha})$ where $\bm \alpha = {\left[{\alpha}_{0}, {\alpha}_{1}, \ldots, {\alpha}_{k}\right[}^{T}$ (which are different from Eq.\eqref{eq:LinearEq_Scaling}).
Therefore, Eq.\eqref{eq:MinErrorNorm_ReconsFlow} can be transformed to the following equation.
%
\begin{equation}
\begin{split}
  \min_{\bm \alpha} {\| {\bm U} {\bm \Sigma} {\bm W}^{T} - {\bm{U V D_{\alpha} T}} \|_F} ^{2} \\
  = \min_{\bm \alpha} {\| {\bm \Sigma} {\bm W}^{T} - {\bm{V D_{\alpha} T}} \|_F} ^{2}
    \label{eq:MinErrorNorm_Reformulated}
\end{split}
\end{equation}
Then, the above minimization problem is equivalent to minimizing the following function.
\begin{equation}
  \min_{\bm \alpha}~J(\bm \alpha) = {\bm \alpha}^{*} {\bm P} {\bm \alpha} - {\bm q}^{*}{\bm \alpha} - {\bm \alpha}^{*}{\bm q} + s
    \label{eq:MinNormFunction}
\end{equation}
where
\begin{subequations}
    
    \begin{equation}
        {\bm P} = ({\bm V}^{*} {\bm V}) \circ (\overline{{\bm T} {\bm T}^{*}})
    \end{equation}
    \begin{equation}
        {\bm q} = \overline{diag({\bm T} {\bm W} {\bm \Sigma}^{*} {\bm V})} 
    \end{equation}
    \begin{equation}
        s = trace({\bm {\Sigma}}^{*} {\bm {\Sigma}})
    \end{equation}
    
    \label{eq:ComponentsMinNormFunction}
\end{subequations}
The optimal amplitude vector~$\bm \alpha$ is obtained as follows, by minimizing the function~$J(\bm \alpha)$.
\begin{equation}
   {\bm \alpha} = {\bm P}^{-1} {\bm q}
    \label{eq:OptimalAmplitude}
\end{equation}

In the Sparsity-Promoting DMD, sparsity structure of amplitude is induced by additional parameter~$\gamma$ with $l_1$-norm of unknown amplitudes~$\alpha_{k}$ which penalize the number of non-zero element in the vector of unknown amplitudes~$\alpha_{k}$. Namely the minimization problem is modified from Eq.\eqref{eq:MinNormFunction} to the following.
%
\begin{equation}
  \min_{\bm \alpha}~J(\bm \alpha)~+~\gamma \sum_{k=0}^{r-1} {|{\alpha}_{k}|}
    \label{eq:MinNorm_SPDMDSP}
\end{equation}
$\gamma$ is the positive regularization parameter which determine the sparsity of the amplitude vector $\bm \alpha$.
It means the stronger emphasis of the values of non-zero elements in amplitude vector $\bm \alpha$ by the larger $\gamma$, which results in smaller numbers of non-zero amplitudes are obtained.

After sparsity matrix is obtained from the above solution, then sparsity structure of the unknown vector of amplitudes and determine only the non–zero amplitude as the solution of the following convex optimization problem. (polishing process)
%
\begin{equation}
\begin{split}
  \min_{\bm \alpha}~J(\bm \alpha)\\
  subject~to~{\bm E}^{T} {\bm \alpha} = 0
    \label{eq:MinNorm_SPDMDPOL}
\end{split}
\end{equation}
In Eq.\eqref{eq:MinNorm_SPDMDPOL}, matrix $\bm E$ represents the sparsity structure of the amplitude vector found by the optimization process of Eq.\eqref{eq:MinNorm_SPDMDSP}.

In this paper, alternating direction method of multipliers (ADMM) algorithm is adopted in this paper to solve convex optimization problem, which is same optimization algorithm adopted in~\cite{Jovanovic2014}.

\section{Incremental Algorithm of Total Dynamic Mode Decomposition}
On-the-fly algorithms of DMD is firstly proposed by Hemati, et al., which are called Streaming DMD (SDMD)~\cite{Hemati2014} and Streaming Total DMD (STDMD)~\cite{Hemati2016}.
These algorithms seem to be very useful especially for huge engineering application such as unsteady vehicle aerodynamics, as DMD computation can be performed in parallel to CFD simulation and requires much less memory than any conventional DMD algorithms.
However, it seems that there is an issue in practical use of Streaming DMD algorithms for the analysis of unsteady flow structures which is the available scaling method to compute DMD amplitudes after Streaming DMD algorithm is only the scaling by the one snapshot vector of the flow field, as the snapshot matrix is not saved on the memory or disk space, if it is performed on-the-fly.
As explained in the section 2.4, scaling of DMD modes by the first snapshot (or a certain snapshot) can sometimes mislead, especially for the identification of the dominant modes in the flow field.

Therefore, we propose the new on-the-fly algorithm of the DMD. This algorithm is achieved by using Incremental Singular Value Decomposition (Incremental SVD)~\cite{Brand2002,Brand2006,Oxberry2017}, instead of Gram-Schmidt orthogonalization used in Streaming DMD and Streaming Total DMD.
To achieve this algorithm, firstly the alternative algorithm of Total DMD (Alternative TDMD) is proposed which can be computed only from singular vectors and values of augmented snapshot matrix.
Afterwards, this algorithm can be performed on-the-fly, if singular vectors and values are computed by incremental SVD, instead of conventional SVD.
This modification of Total DMD algorithm seems to be similar to STDMD algorithm~\cite{Hemati2016}, but there are some advantages in this new algorithm, which are (1) Sparsity-Promoting DMD (SPDMD)~\cite{Jovanovic2014} can be performed after the on-the-fly algorithm (2) There are various choices for the algorithm of incremental SVD.
Especially the combination of on-the-fly algorithm of DMD and SPDMD seems to be very beneficial for the identification of the dominant flow structure in engineering applications, as it can be performed with very small memory and without saving the flow field data on the disk.

In this section, firstly the alternative algorithm of TDMD is derived and explained. This Alternative TDMD algorithm can be performed not only by incremental SVD but also by conventional SVD.
Afterwards, incremental SVD algorithm proposed by Oxbery, et al.~\cite{Oxberry2017} which is adopted in this paper is introduced.  Finally, modified SPDMD algorithm which is performed after the Alternative TDMD is derived and explained.
%
%

\subsection{Alternative Total DMD for achieving incremental DMD algorithm}
As explained in above, if DMD modes are computed by using only singular vectors and values of snapshot matrix of the flow field, the algorithm can be performed on-the-fly when conventional singular value decomposition in the algorithm can be replaced with incremental singular value decomposition algorithm.
Therefore, alternative Total DMD (Alternative TDMD) algorithm is proposed, where DMD modes are computed only from singular vectors and values of augmented snapshot matrix.

Firstly, the singular value decomposition on augmented snapshot matrix is performed as follows.
\begin{equation}
  {\bm Z} = \left[
    \begin{array}{c}
      \bm X \\
      \bm Y
    \end{array}
    \right]
    = {\bm{U_{Z}}} {\bm{\Sigma_{Z}}} {\bm{W_{Z}}}^{T} 
  \label{eq:SVD_Z}
\end{equation}
In the similar process as STDMD by Hemati, et al.~\cite{Hemati2016}, each snapshot matrix $\bm  X \in  \mathbb{R}^{n \times m}$ and $\bm Y \in  \mathbb{R}^{n \times m}$ can be reconstructed from SVD matrices, as shown in the following. 
%
\begin{equation}
  \bm X = \left[ {\bm I}~~{\bm 0} \right] {\bm{{\Sigma}_{Z}}} {\bm{W_{Z}}}^{T}
  \label{eq:SVDZ_ReconsX}
\end{equation}
%
\begin{equation}
  {\bm Y} = \left[ {\bm 0}~~{\bm I} \right] {\bm{{\Sigma}_{Z}}} {\bm{{W}_{Z}}}^{T}
  \label{eq:SVDZ_ReconsY}
\end{equation}
where~${\bm I} \in \mathbb{R}^{n \times n}$ represents identity matrix, and $\left[ {\bm I}~~{\bm 0} \right] \in \mathbb{R}^{n \times 2n}$ is multiplied on~${\bm{U_{Z}}}$ in Eq.\eqref{eq:SVDZ_ReconsX} just for extracting upper~$n$ rows of~${\bm{U_{Z}}}$. In the same manner, $\left[ {\bm 0}~~{\bm I} \right] \in \mathbb{R}^{n \times 2n}$ in Eq.\eqref{eq:SVDZ_ReconsY} extracts lower n rows of ~${\bm{U_{Z}}}$.
Therefore, DMD operator shown in Eq.\eqref{eq:DMDOperator} can be computed by using singular values and vectors, as follows.
%
\begin{equation}
\begin{split}
  {\bm A} = {\bm Y} {\bm X}^{+} &= ( \left[ {\bm 0}~~{\bm I} \right] {\bm{U_{Z}}} {\bm{\Sigma_{Z}}} {\bm{W_{Z}}}^{T} ) {({\bm{W_{Z}}} {\bm{\Sigma_{Z}}}^{-1} \left[ {\bm I}~~{\bm 0} \right]{\bm{U_{Z}}})}^{+}  \\
  &= (\left[ {\bm 0}~~{\bm I} \right] {\bm{U_{Z}}}) (\left[ {\bm I}~~{\bm 0} \right] {\bm{U_{Z}}})^{+}
  \label{eq:altTDMD_DMDoperator}
\end{split}
\end{equation}
As SVD is performed on augmented snapshot matrix $\bm Z$, dimensional consistency between $\bm X$ and $\bm Y$ when constructing DMD operator is conserved, and Total DMD procedure is also included in this algorithm as SVD is applied on $\bm Z$. 

Next, projecting very huge DMD operator matrix onto orthogonal basis is considered to perform the low dimensional approximation of eigendecomposition, following Eq.\eqref{eq:Eq2110} $\sim$ \eqref{eq:Eq2112}. In order to derive the orthogonal basis, additional singular value decomposition is performed on first ~$n$ rows of~$\bm{U_{Z}}$, as follows.
%
\begin{equation}
  (\left[ \bm I~~\bm 0 \right]) = {\bm{U_{X}}} {\bm{\Sigma_{X}}} {\bm{W_{X}}}^{T}
  \label{eq:SVD_Uzx}
\end{equation}
Then Projected DMD operator is computed by projecting DMD operator~$\bm{A}$ in Eq.\eqref{eq:altTDMD_DMDoperator} onto orthogonal basis~${\bm{U_{X}}}$, as follows.
%
\begin{equation}
\begin{split}
  {\bm{\tilde{A}}} = {\bm{U_{X}}}^{T} {\bm A} {\bm{U_{X}}} = {\bm{U_{X}}}^{T} ((\left[ {\bm 0}~~{\bm I} \right] {\bm{U_{Z}}}) {({\bm{U_{X}}} {\bm{\Sigma_{X}}} {\bm{W_{X}}}^{T})}^{+}) {\bm{U_{X}}} \\
  = {\bm{U_{X}}}^{T} (\left[ {\bm 0}~~{\bm I} \right] {\bm{U_{Z}}}) {\bm{W_{X}}} {\bm{\Sigma_{X}}}^{-1}
  \label{eq:altTDMD_projDMDoperator}
\end{split}
\end{equation}
As explained in Eq.\eqref{eq:Eq2110} $\sim$ \eqref{eq:Eq2112}, DMD modes are computed by projecting eigenvectors of~${\bm{\tilde{A}}}$ on orthogonal basis~${\bm{U_{X}}}$, as shown in the following.
%
\begin{align}
  {\bm{\tilde{A} {V}}} &= \bm{{V} \Lambda}
  \label{eq:eig_prjAtilde_altTDMD} \\
%
%
  {\bm{\hat{\Phi}}} &= {\bm{U_{X} V}}
  \label{eq:DMDmode_altTDMD}
\end{align}
If SVD on $\bm Z$ shown in Eq.\eqref{eq:SVD_Z} is performed by incremental SVD, this TDMD algorithm can be performed on-the-fly, as projected DMD operator shown in Eq.\eqref{eq:altTDMD_projDMDoperator} and DMD modes in Eq.\eqref{eq:DMDmode_altTDMD} can be computed after incrementally updating process of SVD matrices.  Therefore, the Eq.\eqref{eq:SVD_Z} is computed by conventional SVD, we call this algorithm ``Alternative TDMD''. On the other hand, if the Eq.\eqref{eq:SVD_Z} is computed by incremental SVD, we call the algorithm ``Incremental TDMD'' in this paper.

\subsection{Incremental Singular Value Decomposition}
Incremental SVD (Incremental SVD) algorithm is the on-the-fly algorithm of SVD~\cite{Brand2002,Brand2006,Oxberry2017}, which compute singular vectors and values incrementally when new snapshot is obtained.
Therefore, Incremental SVD can be performed with smaller memory and in parallel to CFD simulation without saving the flow field data on the disk space.

In this paper, modified version of Brand’s incremental SVD proposed by Oxberry, et al.~\cite{Oxberry2017} is adopted.
It is because this algorithm is easy to be implemented, therefore, it seems to be suit for the first implementation to investigate the applicability the of Incremental TDMD algorithm.
In this section, we introduce the modified incremental SVD algorithm is briefly introduced, according to~\cite{Brand2002,Brand2006,Oxberry2017}, but the algorithm with the single-column updates is adopted, as presented in~\cite{Oxberry2017}.

Firstly, existing rank-r SVD of a matrix $\bm Z$ is defined as follows as an inputs to the incremental SVD algorithm.
%
\begin{equation}
  {\bm Z} = {\bm{U_{Z}}} {\bm{\Sigma_{Z}}} {\bm{{W_{Z}}}}^{T} + {\bm R}
  \label{eq:SVDZ_wError}
\end{equation}
where ${\bm{U_{Z}}} {\bm{\Sigma_{Z}}} {\bm{{W_{Z}}}}^{T}$ is equivalent to Eq.\eqref{eq:SVD_Z} but rank-r truncated SVD matrices of $\bm Z$, and $\bm R$ is resultant error by truncation. 
When new augmented snapshot (column) vector~${\bm z}_{i} = \left[ {\bm x}_{i}~;~{\bm y}_{i}\right]$ is obtained, rank-r truncated SVD matrices is updated.
This incremental SVD algorithm is derived from the following identity.
%
\begin{equation}
\begin{split}
      \left[ {\bm Z} ~~ {\bm z}_{i} \right]
      &= \left[ {\bm{U_{Z}}} {\bm{{\Sigma}_{Z}}} {\bm{W_{Z}}}^{T} ~~ {\bm z}_{i} \right]
      \\
      &= \left[ {\bm{U_{Z}}}   ({\bm I} - {\bm{U_{Z}}} {\bm{U_{Z}}}^{T}) {{\bm z}_{i}}/p \right]
            \left[
                \begin{array}{cc}
                    {\bm{{\Sigma}_{Z}}} & {\bm{U_{Z}}}^{T} {\bm z}_{i} \\
                    {\bm 0}  & p
                \end{array}
            \right]
            \left[
                \begin{array}{cc}
                    {\bm{W_{Z}}} &{\bm 0} \\
                    {\bm 0}   & 1
                \end{array}
            \right]^{T}
    \\
        &=
        \left[ {\bm{U_{Z}}}~~{\bm q} \right]
        \left[
            \begin{array}{cc}
                {\bm{{\Sigma}_{Z}}} & {\bm l} \\
                {\bm 0}  & p
            \end{array}
        \right]
        \left[
            \begin{array}{cc}
                {\bm{W_{Z}}} &{\bm 0} \\
               {\bm 0}   &1
            \end{array}
         \right]^{T}
\end{split}
          \label{eq:iSVD_append}
\end{equation}
where $p = \|( {\bm I} - {\bm{U_{Z}}} {\bm{U_{Z}}}^{T}) {\bm z}_{i} \|$, ${\bm q} = ({\bm I} - {\bm{U_{Z}}} {\bm{U_{Z}}}^{T}) {\bm z}_{i}/p$ and ${\bm l} = {\bm{U_{Z}}}^{T} {\bm z}_{i}$.
Here, $p$ is the length of the orthogonal projection of ${\bm z}_{i}$ onto the subspace of orthogonal basis ${\bm{U_{Z}}}$. 
The vector ${\bm q}$ is the normalized, orthogonal projection of ${\bm z}_{i}$ onto the subspace which is orthogonal to ${\bm{U_{Z}}}$.
${\bm l}$ is the projection of ${\bm z}_{i}$ onto the basis of ${\bm{U_{Z}}}$.

Additional SVD is performed on the middle matrix in Eq.\eqref{eq:iSVD_append} to obtain new diagonal singular value matrix, as shown in the following.
%
\begin{equation}
    \begin{bmatrix}
        {\bm{{\Sigma}_{Z}}} &{\bm l} \\
        {\bm 0}   &p
    \end{bmatrix}
    = {\bm{U_{SZ}}} {\bm{{\Sigma}_{SZ}}} {\bm{W_{SZ}}}^{T}
    \label{eq:iSVD_additionalSVD}
\end{equation}
Then, updated SVD on $\left[ {\bm Z}~~{\bm z}_{i} \right]$ is described as follows.
%
\begin{equation}
\begin{split}
    \left[ {\bm Z} ~~{\bm z}_{i} \right]
    &= \left[ {\bm{U_{Z}}} ~~{\bm q} \right] {\bm{U_{SZ}}} {\bm{{\Sigma}_{SZ}}} {\bm{{W}_{SZ}}}^{T}
        {\begin{bmatrix}
            {\bm{{W}_{Z}}} &{\bm 0} \\
            {\bm 0}   &1    
        \end{bmatrix}}^{T} \\
    &= {\bm{U_{Z.updated}}} {\bm{{\Sigma}_{Z.updated}}} {\bm{{W}_{Z.updated}}}^{T}
    \label{eq:iSVD_update}
\end{split}
\end{equation}
where
%
\begin{subequations}
    %
    \begin{align}    
        {\bm{U_{Z.updated}}}~&\leftarrow~ \left[ {\bm{U_{Z}}} ~~{\bm q} \right] {\bm{U_{SZ}}} \\
        {\bm{{\Sigma}_{Z.updated}}}~&\leftarrow~{\bm{{\Sigma}_{SZ}}} \\
        {\bm{{W}_{Z.updated}}}~&\leftarrow~
            \left[
            \begin{array}{cc}
                {\bm{W_{Z}}} &{\bm 0} \\
                {\bm 0}   &1    
            \end{array}
            \right]
            {\bm{W_{SZ}}} 
    \end{align}
\end{subequations}
Then, the rank of matrices is updated from $r$ to $r+1$.

Additionally, automatic truncation is implemented in~\cite{Brand2002,Oxberry2017}.  If $p < {\epsilon}_{SVD}$ for a small ${\epsilon}_{SVD}$ near the limits of machine precision, then vector ${\bm q}$ must have zero norm, and the rank of singular vectors are not increased. 
Therefore, the updated SVD matrices are replaced as follows.
%
\begin{subequations}
    %
    \begin{align}
        {\bm {U_{Z.updated}}}~&\leftarrow~ {\bm{U_{Z}}} {\bm{U_{SZ}}}_{1:r,1:r} \\
        {\bm{{\Sigma}_{Z.updated}}}~&\leftarrow~{\bm{{\Sigma}_{SZ}}}_{1:r,1:r} \\
        {\bm{{W}_{Z.updated}}}~&\leftarrow~ {\bm{{W}_{Z}}}  {\bm{{W}_{SZ}}}_{:,1:r}
    \end{align}
\end{subequations}

In this paper, another truncation of singular vectors and values are performed according to the prescribed maximum rank~$r_{max}$ at the last of one updating process, when~$r_{updated}~>~r_{max}$, as follows.
%
\begin{subequations}
    %
    \begin{align}
        {\bm{U_{Z.truncated}}}~&\leftarrow~ {\bm{U_{Z.updated}}}_{:,1:r} \\
        {\bm{{\Sigma}_{Z.truncated}}}~&\leftarrow~ {\bm{{\Sigma}_{Z.updated}}}_{1:r,1:r} \\
        {\bm{W_{Z.truncated}}}~&\leftarrow~ {\bm{W_{Z.updated}}}_{:,1:r}
    \end{align}
\end{subequations}

As incremental SVD algorithm introduced here is updated by using single column vector~${\bm z}_{i}$, SVDmatrices are firstly initialized by using the first snapshot~${\bm z}_{0}$, as the following
%
\begin{subequations}
    %
    \begin{align}
        {\bm{U_{Z.init}}}~&\leftarrow~ {{\bm z}_{0}} / \|{{\bm z}_{0}}\| \\
        {\bm{{\Sigma}_{Z.init}}}~&\leftarrow~ \|{{\bm z}_{0}}\| \\
        {\bm{W_{Z.init}}}~&\leftarrow~ 1 
    \end{align}
\end{subequations}
In this paper, we adopted incremental SVD algorithm to compute singular vectors and values of augmented snapshot matrix $\bm Z$ on-the-fly, to replace conventional SVD in Eq.\eqref{eq:SVD_Z} with incremental SVD.
%

\subsection{Sparsity-promoting DMD after Alternative Total DMD}
One of advantages of Alternative TDMD algorithm is that Sparsity-Promoting DMD (SPDMD) can be performed even if TDMD is performed on-the-fly by using incremental SVD.
It is because low-rank singular vectors and values of augmented snapshot matrix $\bm Z$ are conserved after the conventional SVD or incremental SVD, which enable the reduced-rank approximation of the snapshot matrix by the singular vectors and values.
Therefore, it can be investigated the error norm between reconstructed snapshot matrix from singular vectors and values, and reconstructed snapshots from DMD modes, which result in the optimal amplitude of DMD modes and the capability of SPDMD.
As described in Eq.\eqref{eq:MinErrorNorm_ReconsFlow}, the error norm shown in the following is minimized to compute optimal amplitude.
%
\begin{equation}
  \min_{\bm{\alpha}} {\|{\bm X} - {\bm{\hat{\Phi}}} {\bm{D_{\alpha}}} {\bm T}\|_F} ^{2}
    \label{eq:MinErrNorm_altTDMD}
\end{equation}
$\bm X$ can be approximated by using low-rank singular values and vectors computed in Eq.\eqref{eq:SVDZ_ReconsX} and Eq.\eqref{eq:SVD_Uzx}, and DMD modes~${\bm{\hat{\Phi}}}$ are computed as Eq.\eqref{eq:DMDmode_altTDMD}, then Eq.\eqref{eq:MinErrNorm_altTDMD} can be reformulated as follows.
%
\begin{equation}
\begin{split}
  &\min_{\bm{\alpha}} {\| \left[ {\bm I} ~~ {\bm 0} \right] {\bm{U_{Z}}} {\bm{{\Sigma}_{Z}}} {\bm{W_{Z}}}^{T}  - {\bm{\hat{\Phi}}} {\bm{D_{\alpha}}} {\bm T}\|_F} ^{2} \\
  = &\min_{\bm{\alpha}} {\| {\bm{U_{X}}} {\bm{{\Sigma}_{X}}} {\bm{W_{X}}}^{T} {\bm{\Sigma}_{Z}} {\bm{W_{Z}}}^{T}  - {\bm{U_{X}}} {\bm V} {\bm{D_{\alpha}}} {\bm T}\|_F} ^{2} \\
  = &\min_{{\bm \alpha}} {\| {\bm{{\Sigma}_{X}}} {\bm{W_{X}}}^{T} {\bm{{\Sigma}_{Z}}} {\bm{W_{Z}}}^{T}  - {\bm V} {\bm{D_{\alpha}}} {\bm T}\|_F} ^{2}
    \label{eq:MinErrNorm_altTDMD_Reformulated}
\end{split}
\end{equation}
Therefore, according to Eq.\eqref{eq:MinNormFunction} and Eq.\eqref{eq:ComponentsMinNormFunction}, the equations for convex optimization in SPDMD is derived as the following.
%
\begin{equation}
  \min_{\bm{\alpha}}~J(\bm{\alpha}) = {\bm{\alpha}}^{*} {\bm{P}} {\bm{\alpha}} - {\bm{q}}^{*}{\bm{\alpha}} - {\bm{\alpha}}^{*}{\bm{q}} + s
    \label{eq:MinNormFunction_altTDMD}
\end{equation}
where
%
\begin{subequations}
    
    \begin{align}
        {\bm P} &= ({\bm V}^{*} ~{\bm V}) \circ (\overline{{\bm T} ~{\bm T}^{*}}) \\
        {\bm q} &= \overline{diag({\bm T} {\bm{W_{Z}}} {\bm{{\Sigma}_{Z}}}^{*} {\bm{W_{X}}} {\bm{{\Sigma}_{X}}}^{*} {\bm V})} \\
        s &= trace({\bm {{\Sigma}_{Z}}}^{*} {\bm{W_{X}}} {\bm{{\Sigma}_{X}}}^{*} {\bm{{\Sigma}_{X}}} {\bm{W_{X}}}^{*} {\bm{{\Sigma}_{Z}}})
    \end{align}
    
    \label{eq:ComponentsMinNormFunction_altDMD}
\end{subequations}
Then, optimal amplitudes and SPDMD amplitudes can be approximated in the same manner as Eq.\eqref{eq:OptimalAmplitude}~$\sim$~\eqref{eq:MinNorm_SPDMDPOL} in the section 2.4.

Therefore, all of matrices used in this SPDMD equations are computed by Alternative TDMD algorithm in the section 3.1., even though SVD in Eq.\eqref{eq:SVD_Z} is performed by using incremental SVD. Therefore, SPDMD amplitudes can be approximated even after Incremental TDMD algorithm.

\section{Flow simulation - Infinite square cylinder}
As a test flow field for the DMD algorithms, the flow around infinite square cylinder is simulated. 
The numerical simulation is performed using OpenFOAM (version 2.2.2) which is open source CFD code developed by openfoam.org. 
The result of the flow simulation is validated by comparson with previous setups~\cite{Bouris1999,Bosch1998}.

\subsection{Computational Setups}
The flow simulation around the infinite square cylinder with $Re = 22,000$ of Reynolds number regarding the length of the square cylinder and inlet velocity is performed according to ~\cite{Bouris1999,Bosch1998}.
The computational grid and domain is shown in Fig.4.1. The size of computational domain and location of the square cylinder is same as the setup in~\cite{Bouris1999}, with respect to the length of the square cylinder. 
Numbers of cartesian grid used for the simulation is~$[224 \times 300]$, and the height of the closest cells to the square cylinder is ${\delta}y$/$h$~=~$6.25 \times 10^{-3}$ which is slightly larger than presented in~\cite{Bouris1999,Bosch1998}. 
However, it is judged that this setup seems to result in the enough relevancy of the simulation for the purpose of the evaluation of DMD algorithms.  
In the simulation,~$D = 1$ m is chosen as the length of the square cylinder, and uniform velocity~$U_{inlet} = 11$ m/sec is implemented as the inlet velocity condition from the slice located in front of the domain, with kinetic viscosity~$\nu = 5 \times 10^{-4}$ reproducing $Re = 22,000$.
This flow simulation is conducted by Large Eddy Simulation with Standard Smagorinsky model with the Smagorinsky constant~$C_s = 0.168$ and the van-driest wall damping function is implemented near the wall.
The flow simulation of infinite square cylinder by OpenFOAM is artificially reproduced by the specific boundary condition on the top and bottom walls with respect to the xy-slice where z-components in the equation are not solved.
As a velocity-pressure coupling algorithm, pimple algorithm is adopted in this simulation.
%
%

\begin{figure}[htbp]
     \begin{tabular}{cc}
       \begin{minipage}[b]{0.5\hsize}
         \centering
         \includegraphics[keepaspectratio, scale=0.25]{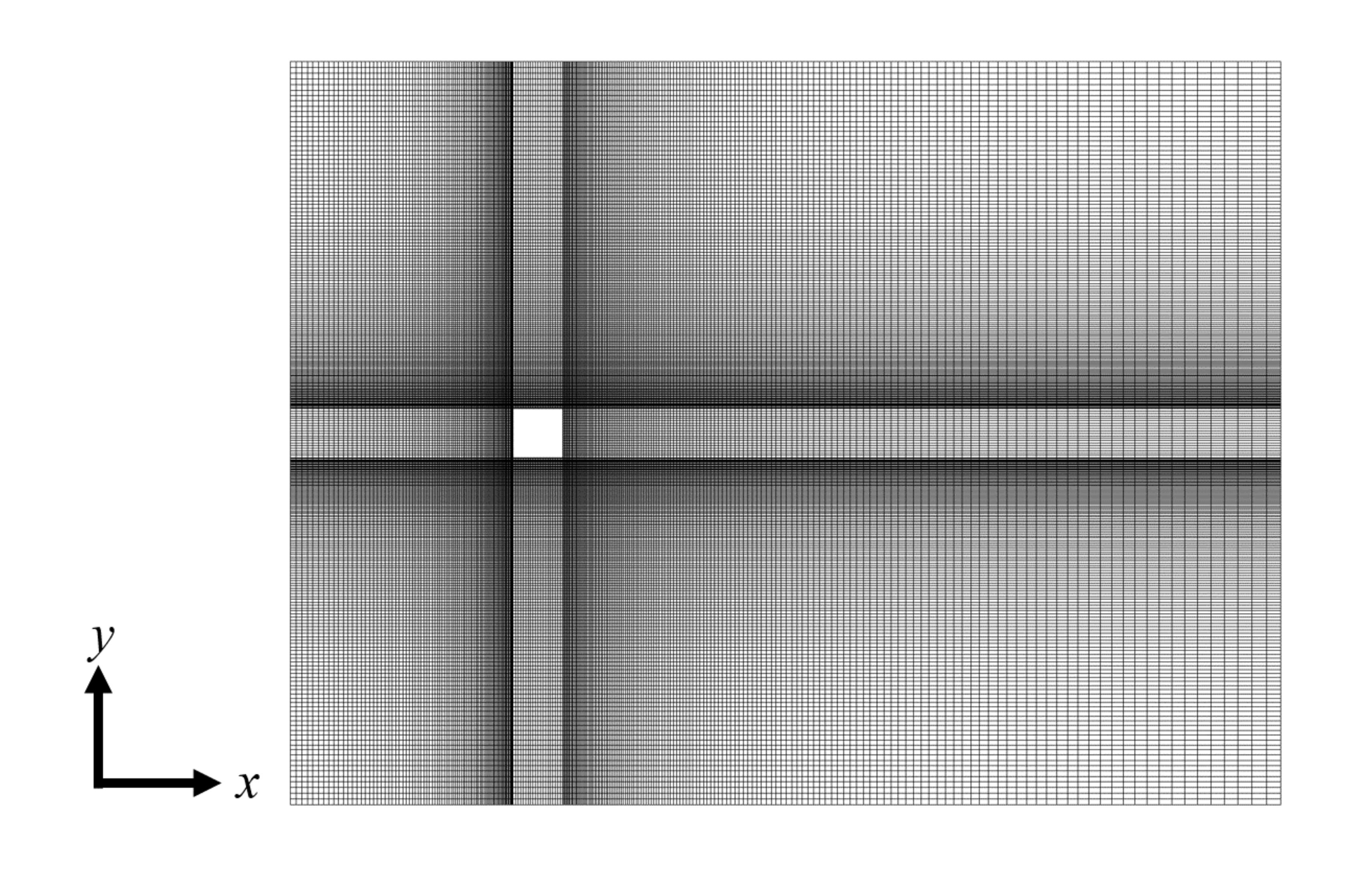}
         \subfloat{(a)~Whole~computational~domain}
       \end{minipage} &
       \begin{minipage}[b]{0.5\hsize}
         \centering
         \mbox{\raisebox{6mm}{\includegraphics[keepaspectratio, scale=0.3]{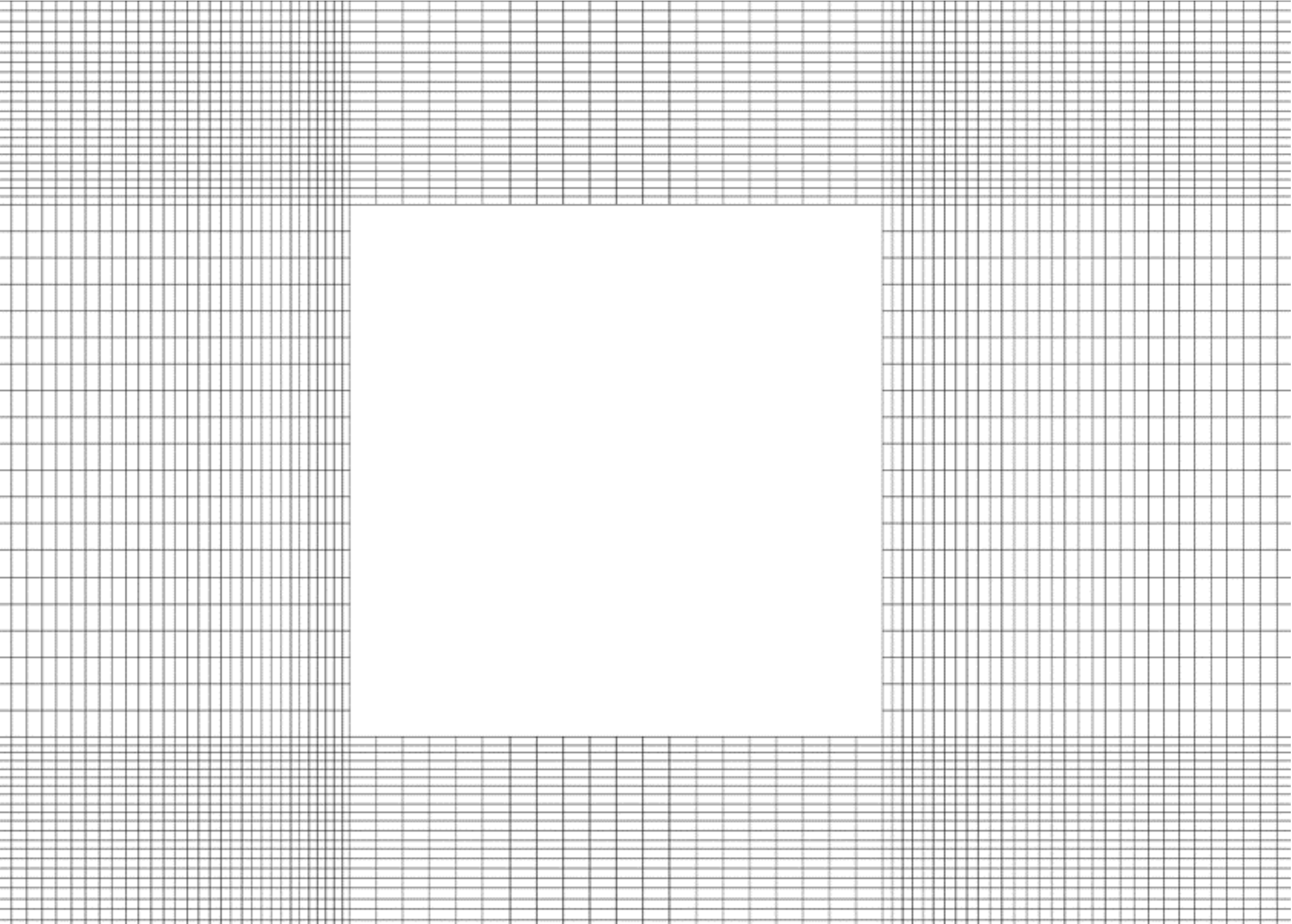}}}
          \subfloat{(b)~Close~up~the~square~cylinder}
       \end{minipage}
     \end{tabular}
         \caption{Computational grid}
         \label{ComputationalGrids}
   \end{figure}

\subsection{The Result of the Numerical Simulation}
For the validation, results from this simulation is compared with the previous setups~\cite{Bouris1999,Bosch1998}.
Firstly the normalized, time averaged streamwise velocity which is measured along the center line with respect to y direction is shown in Fig.4.2. with the previous results presented in~\cite{Bouris1999} where other simulation and experimental results presented in~\cite{Breuer1996, Rodi1993, Lyn1995, Durao1988} are also plotted.
With respect to the time averaged velocity distribution, our simulation seems to have enough relevancy compared to the simulation in~\cite{Bouris1999}.
In addition, the normalized, mean kinetic fluctuation energy~$k = \frac{{\overline{{u^{\prime}}^{2}}} + {\overline{{v^{\prime}}^{2}}}}{2}$ measured in this simulation is shown in Fig.4.3 with the results presented in~\cite{Bouris1999} where another experimental result presented in~\cite{Franke1991} is also plotted.
In consequence, kinetic fluctuation energy seems to be in good agreement with the result presented in~\cite{Bouris1999}.
Therefore, even though there are still some discrepancies with respect to velocity and kinetic fluctuation energy compared to experimental results and previous simulations, however, it seems that relevancy of this simulation is enough for the discussion of the relevancy of DMD algorithms. 
%
%

\begin{figure}[htbp]
    \begin{tabular}{cc}
      \begin{minipage}[t]{0.5\hsize}
        \centering
        \includegraphics[keepaspectratio, scale=0.34]{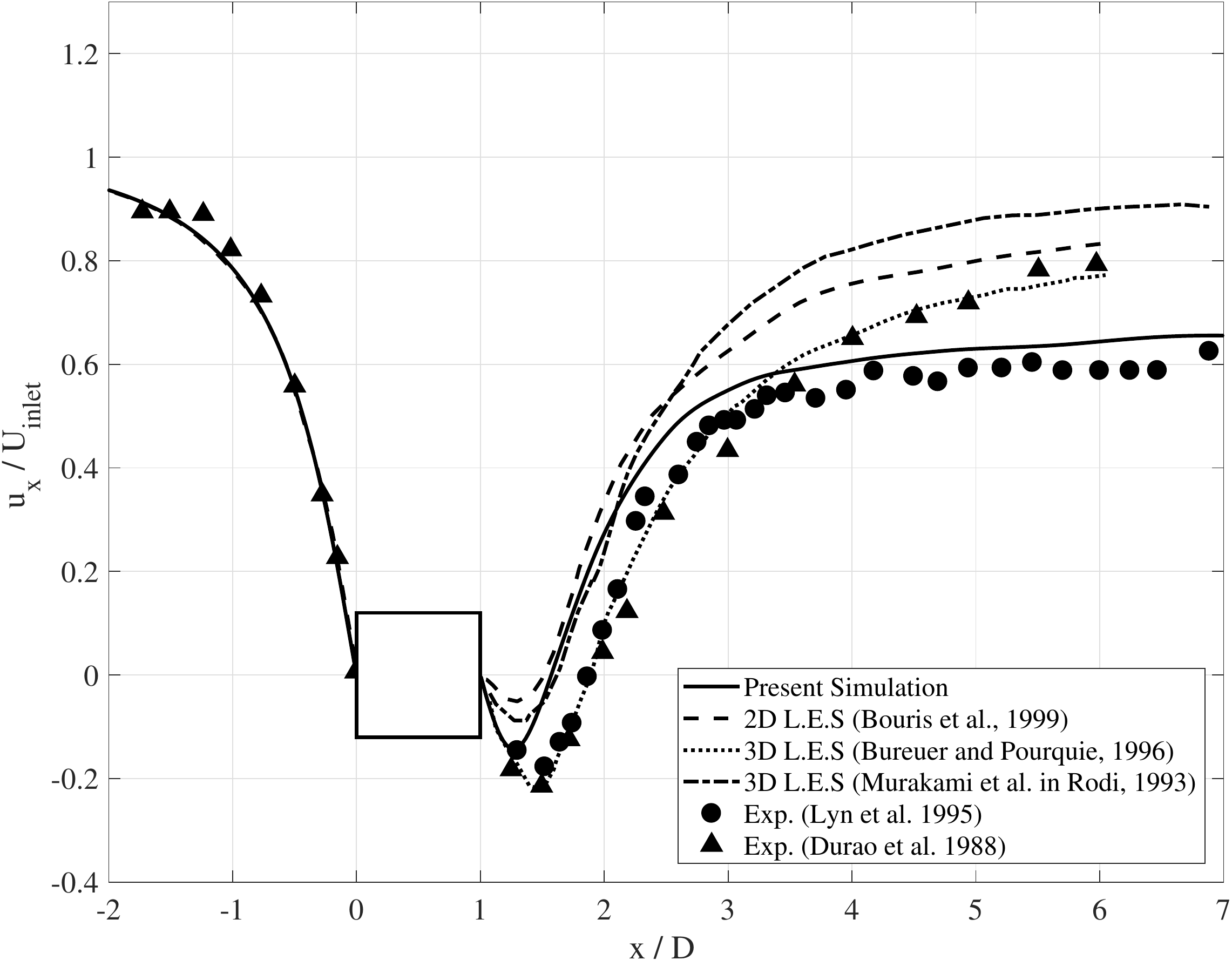}
        \caption{Normalized mean streamwise velocity in the present simulation and presented in~\cite{Bouris1999}}
        \label{UmeanValue}
      \end{minipage} &
      \begin{minipage}[t]{0.5\hsize}
        \centering
        \includegraphics[keepaspectratio, scale=0.34]{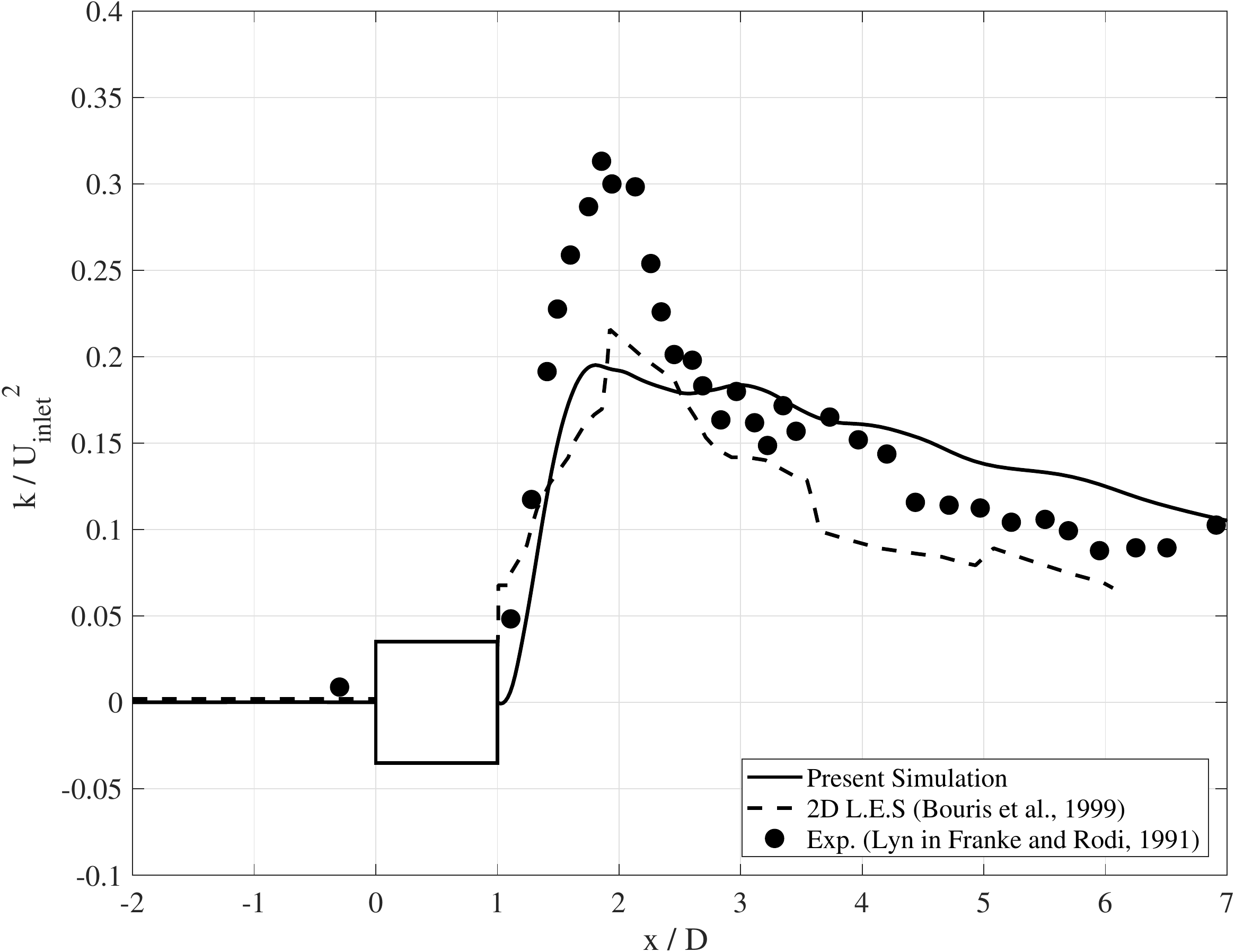}
        \caption{Normalized mean kinetic energy in the present simulation and presented in~\cite{Bouris1999}}
        \label{Uprime2MeanValue}
      \end{minipage}
    \end{tabular}
  \end{figure}
%
%

\section{Results}
In this section, various TDMD algorithms explained in the section 2 and 3 are performed and compared in order to investigate the relevancy of Alternative TDMD algorithm and Incremental TDMD algorithm. 
Every TDMD in this paper are applied on the velocity magnitude around the square cylinder simulated in the section 4.
The values of velocity magnitude are stored in the centers of computational cells across the whole computational domain, which is chosen as a snapshot vector. Numbers of snapshot vectors used for TDMD computations are 1000 snapshots sampled with the time interval~${\Delta}t_\text{DMD} = 5 \times 10^{-3}$ sec.

First, results of conventional TDMD and Alternative TDMD with conventional SVD (in section 3.1) are compared, in order to validate the relevancy of the formulations of Alternative TDMD.
In addition, SPDMD is performed after each TDMD computation to identify dominant modes to be compared.
Second, Incremental TDMD is performed and compared with the Alternative TDMD, where the difference between those algorithms is only the choice of SVD algorithm, in order to investigate the relevancy of Incremental TDMD algorithm.
After computing DMD modes by each Total DMD algorithm, Sparsity-Promoting DMD is also performed respectively to identify dominant modes in the results from each algorithm. 
Please note that every amplitude of DMD (which are by the first snapshot, optimal amplitude and SPDMD amplitude) shown in this section are normalized by the maximum amplitude in each computation.

\subsection{Validation of Alternative TDMD}
Conventional TDMD and Alternative TDMD are applied on the velocity magnitude around the square cylinder.
The maximum rank of POD modes is set to $r_{max} = 107$ out of 999 resultant modes of POD, where the POD modes which are contributing less than 0.01\% with respect to the total variance of singular values of augmented snapshot matrix are truncated.

First, the DMD amplitudes computed by the scaling with the first snapshot of flow field reconstructed from POD modes (explained in the section 2.3) are shown in Fig.5.1. with the Strouhal number of each mode.
Comparing amplitude shown in Fig5.1.(a) and (b), the difference in the amplitude distribution between conventional TDMD and Alternative TDMD is not appeared.
%
%

\begin{figure}[htbp]
     \begin{tabular}{cc}
       \begin{minipage}[b]{0.5\hsize}
         \centering
        \includegraphics[keepaspectratio, scale=0.34]{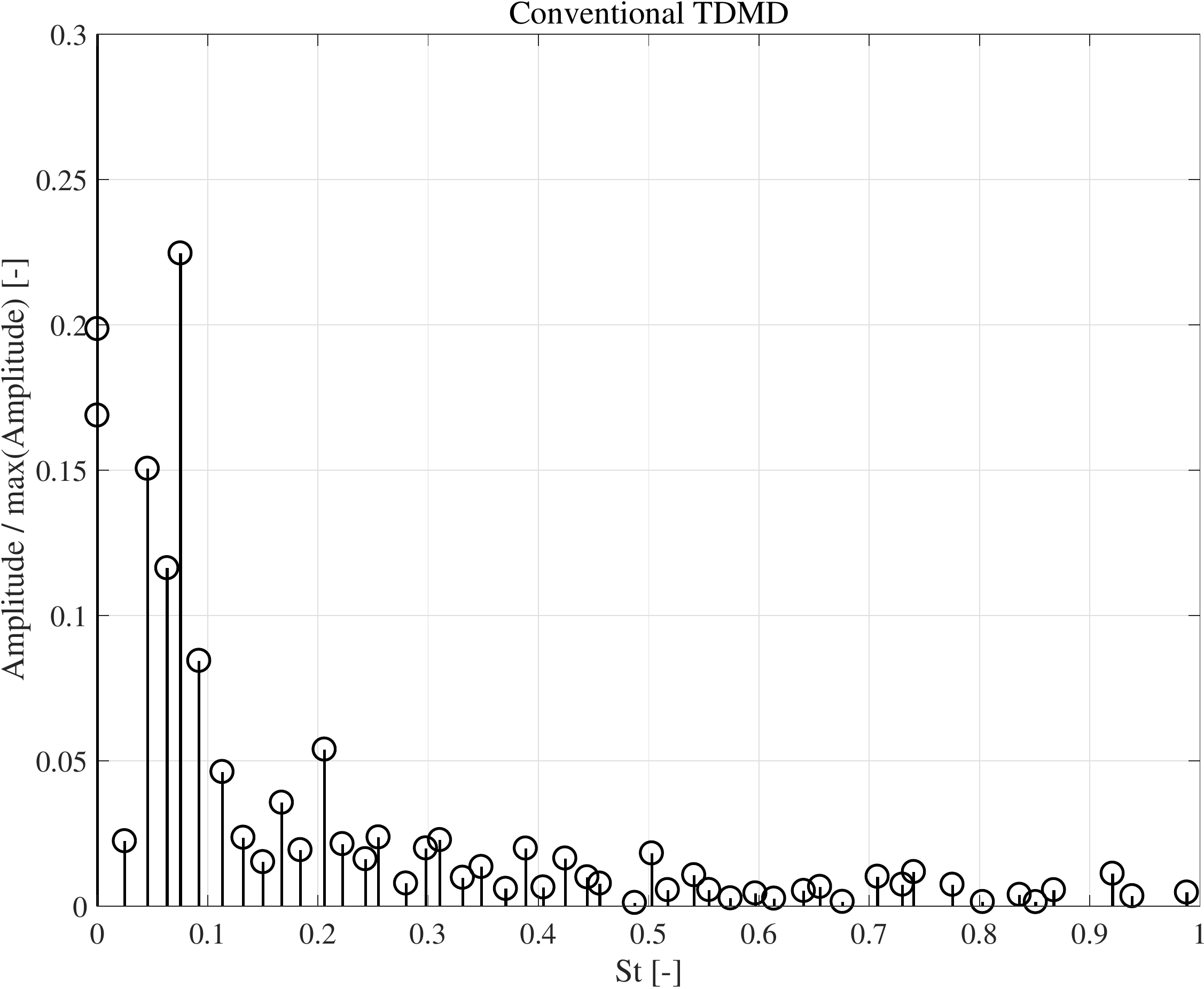}
        \subfloat{(a)~Conventional~TDMD}
       \end{minipage} &
       \begin{minipage}[b]{0.5\hsize}
         \centering
        \includegraphics[keepaspectratio, scale=0.34]{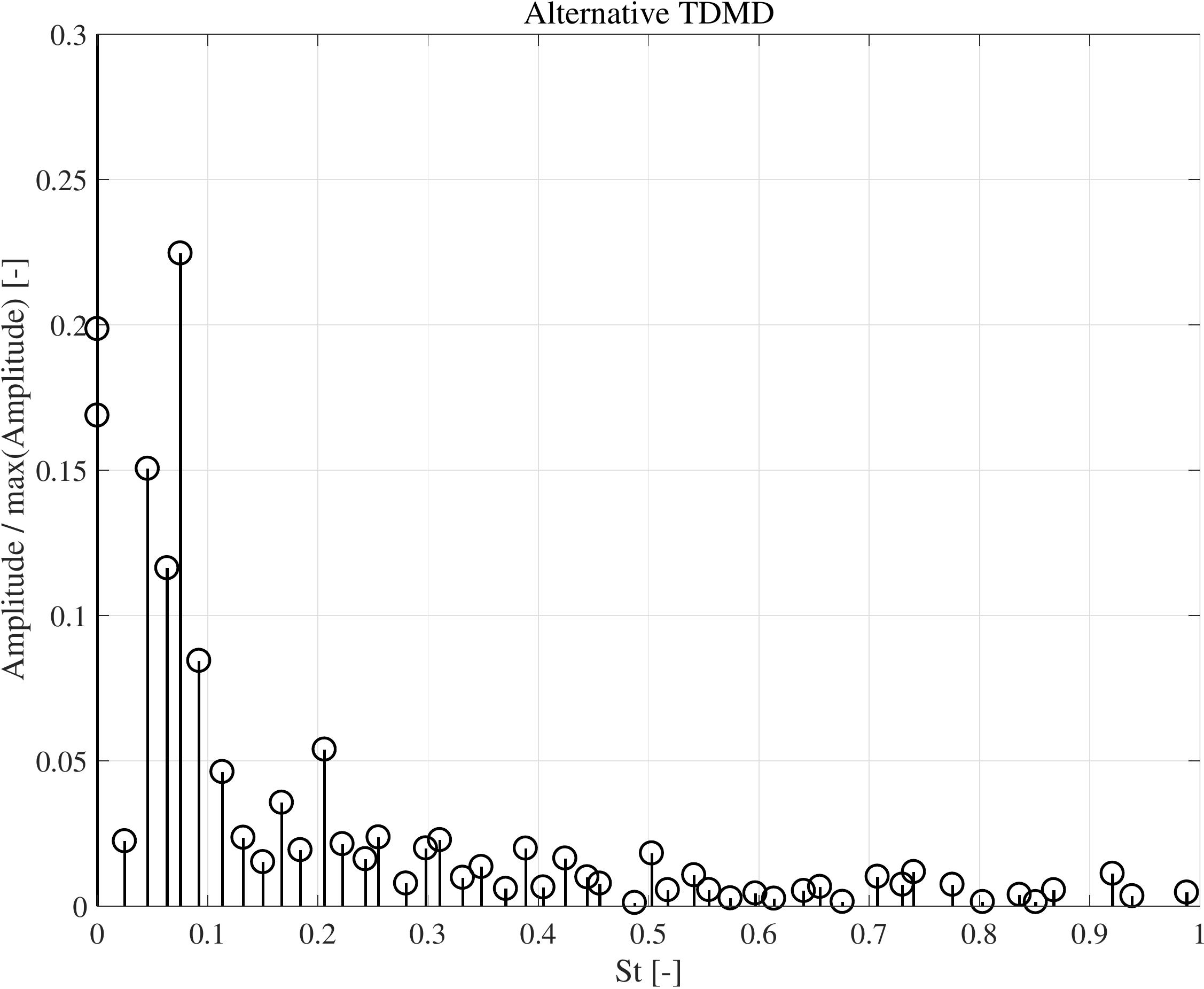}
        \subfloat{(b)~Alternative~TDMD}
       \end{minipage}
     \end{tabular}
         \caption{DMD~amplitude~by~the~first~snapshot~vector}
         \label{DMDmodeAmp_cTDMDvsAltTDMD}
\end{figure}

Second, SPDMD is performed after those TDMD computations. The SPDMD amplitudes are shown in Fig.5.2. with the optimal amplitudes (represented as Optimal amplitude) which is computed by Eq.~\eqref{eq:OptimalAmplitude} without the penalty parameter~$\gamma$.
In Fig.5.2, SPDMD amplitudes having different numbers of non-zero amplitudes are shown. Please note that the numbers of modes having non-zero amplitudes are counted including negative frequency modes.
From Fig.5.2., the difference in SPDMD amplitude between conventional TDMD and Alternative TDMD is not observed, even if the different numbers of mode having non-zero amplitude are remained. 
%
%

\begin{figure}[htbp]
     \begin{tabular}{cc}
       \begin{minipage}[b]{0.5\hsize}
         \centering
        \includegraphics[keepaspectratio, scale=0.34]{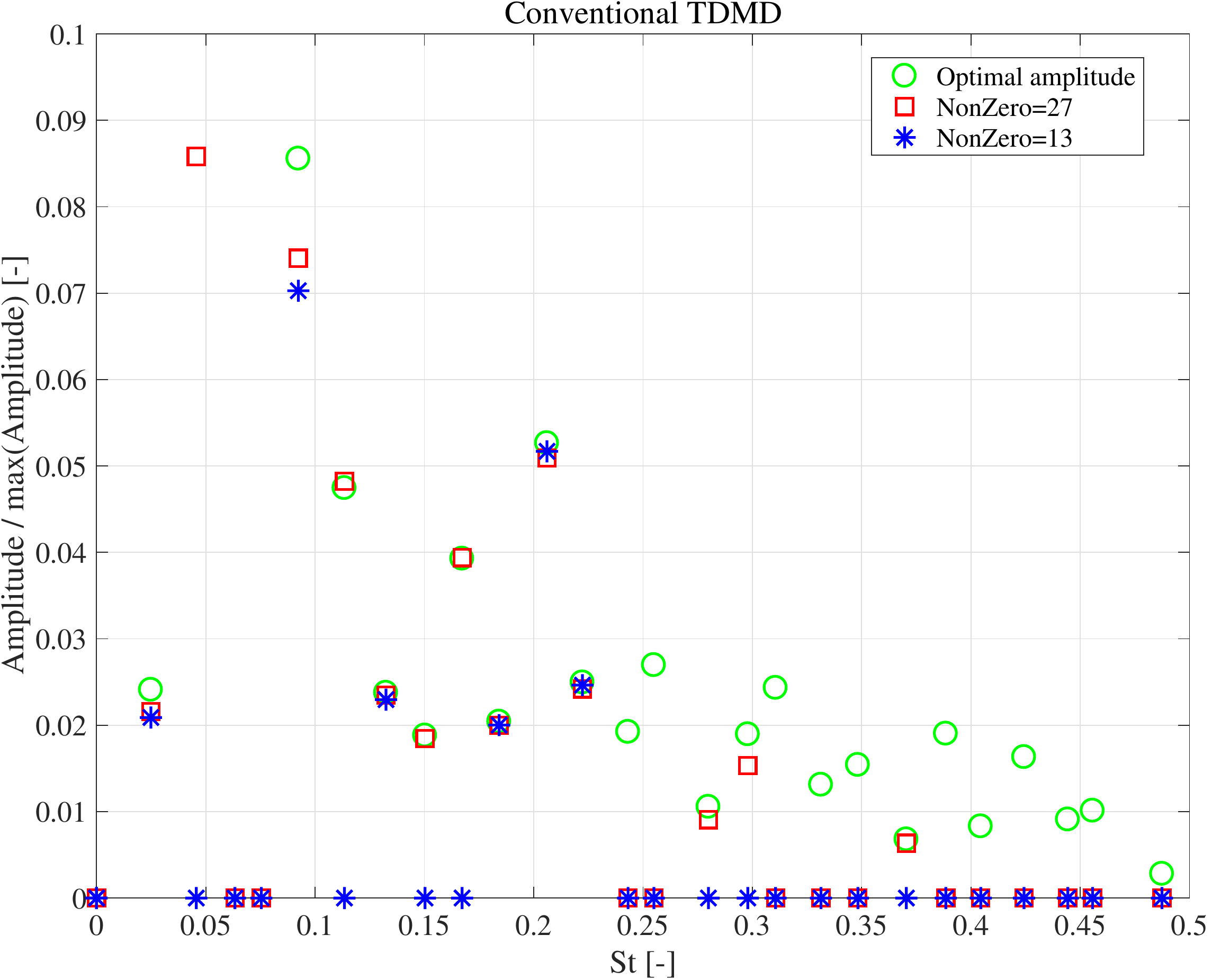}
        \subfloat{(a)~Conventional~TDMD}
       \end{minipage} &
       \begin{minipage}[b]{0.5\hsize}
         \centering
        \includegraphics[keepaspectratio, scale=0.34]{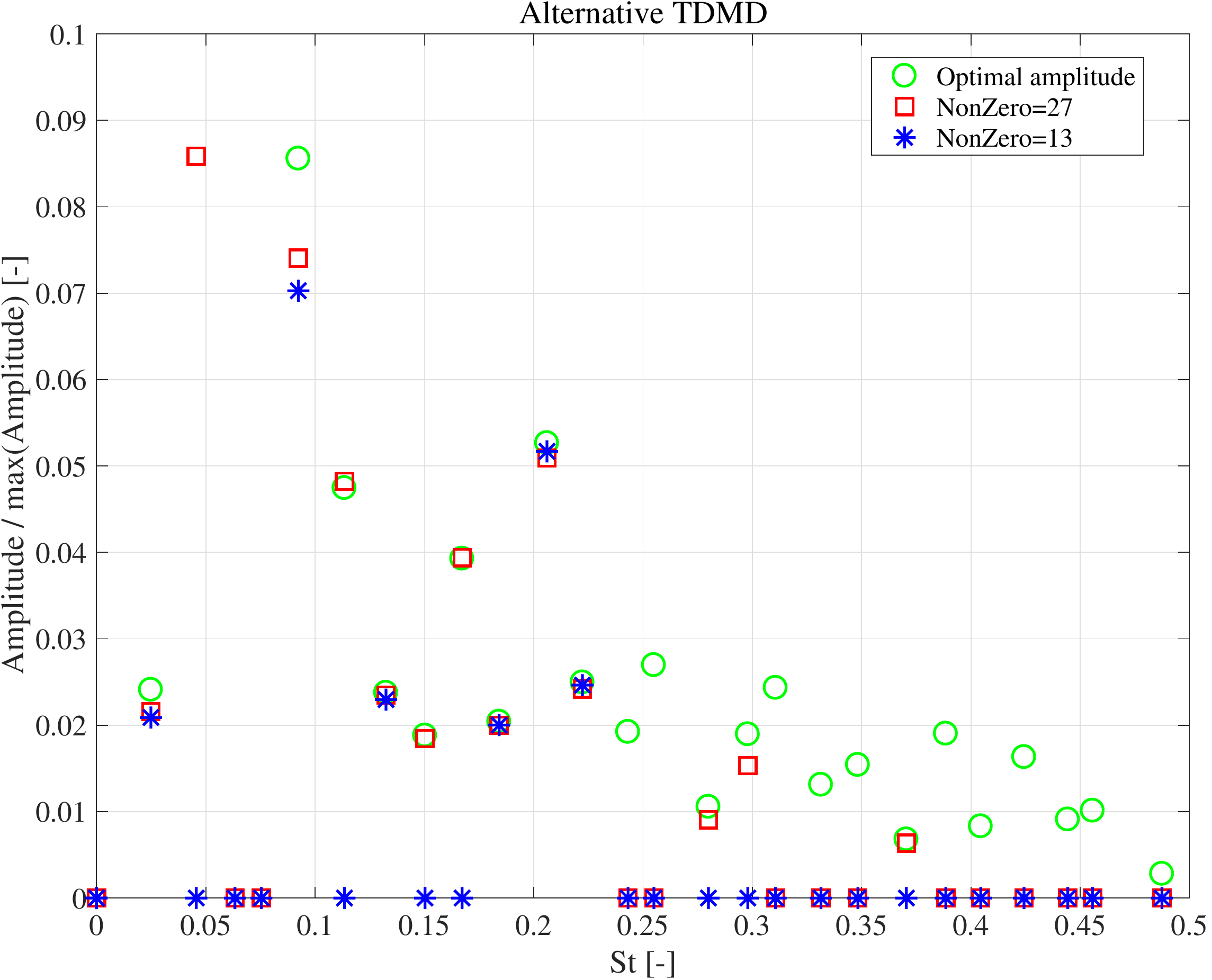}
        \subfloat{(b)~Alternative~TDMD}
       \end{minipage}
     \end{tabular}
         \caption{SPDMD~amplitudes~and~the~optimal~amplitude}
         \label{SPDMDmodeAmp_cTDMDvsAltTDMD}
\end{figure}

Finally, the spatial distributions of dominant DMD mode are examined. Fig.5.3. shows SPDMD amplitude distributions when remained non-zero amplitudes are 27, which is same as the amplitude distribution described as red square marker in Fig.5.2.
The spatial distribution of dominant DMD modes, with respect to the SPDMD amplitude shown in Fig.5.3, located at~$St = 0.0456,~0.0923,~0.1134~\text{and}~0.2061$ are visualized in Fig.5.4 for the comparison.
According to Fig.5.4, no difference is observed in the spatial distribution at the same frequency between the result by the conventional TDMD and Incremental TDMD.

Hence, it is concluded that the Alternative TDMD algorithm combined with SPDMD results in the same as the conventional TDMD combined with SPDMD.
%
%

\begin{figure}[htbp]
     \begin{tabular}{cc}
       \begin{minipage}[b]{0.5\hsize}
         \centering
        \includegraphics[keepaspectratio, scale=0.42]{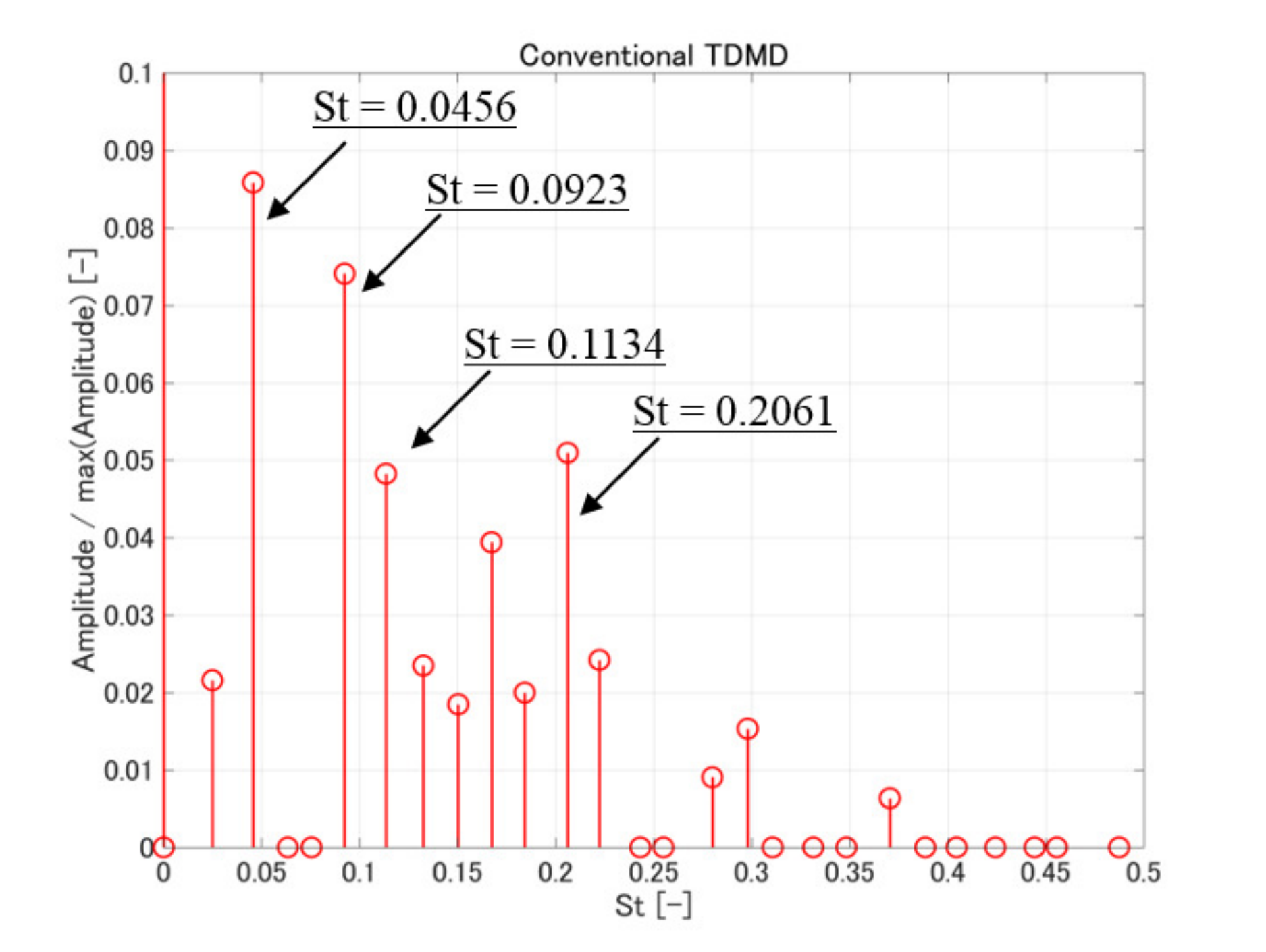}
        \subfloat{(a)~Conventional~TDMD}
       \end{minipage} &
       \begin{minipage}[b]{0.5\hsize}
         \centering
        \includegraphics[keepaspectratio, scale=0.42]{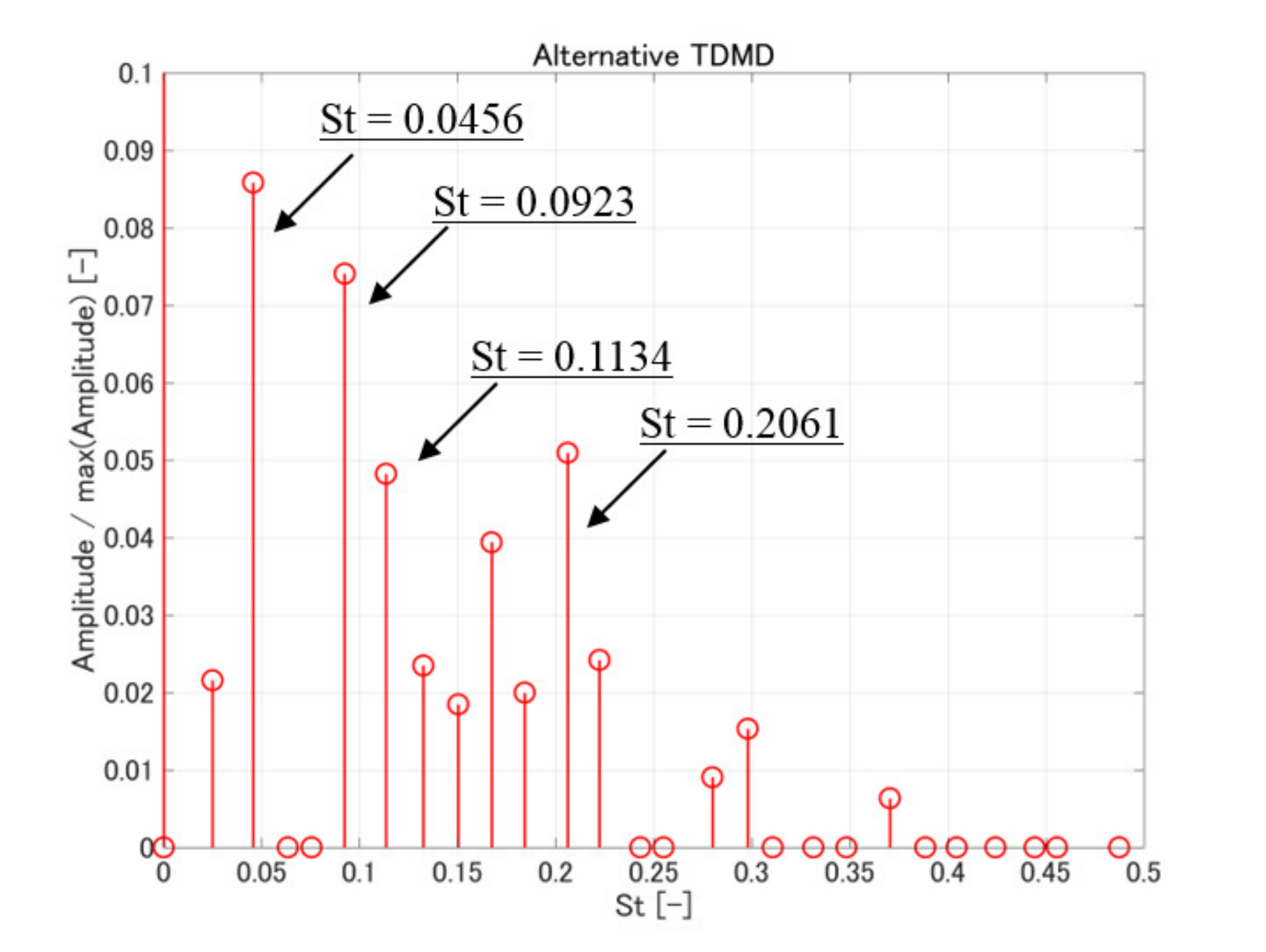}
        \subfloat{(b)~Alternative~TDMD}
       \end{minipage}
     \end{tabular}
         \caption{SPDMD amplitude (27 modes having non-zero amplitude)}
         \label{SPDMDmodeAmp_cTDMDvsAltTDMD_27modes}
\end{figure}
%

%
\begin{figure}[htbp]
    \begin{center}
             \begin{tabular}{cccc}
               \begin{minipage}[b]{0.25\hsize}
                 \centering
                \includegraphics[keepaspectratio, scale=3]{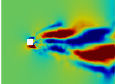}\hfill
                \subfloat{(a.1)~St~=~0.0456}
               \end{minipage} \hfill
               \begin{minipage}[b]{0.25\hsize}
                 \centering
                \includegraphics[keepaspectratio, scale=3]{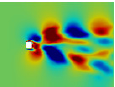}\hfill
                \subfloat{(a.2)~St~=~0.0923}
               \end{minipage} \hfill
               \begin{minipage}[b]{0.25\hsize}
                 \centering
                \includegraphics[keepaspectratio, scale=3]{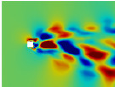}\hfill
                \subfloat{(a.3)~St~=~0.1134}
               \end{minipage} \hfill
               \begin{minipage}[b]{0.25\hsize}
                 \centering
                \includegraphics[keepaspectratio, scale=3]{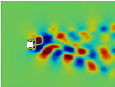}\hfill
                \subfloat{(a.4)~St~=~0.2061}
               \end{minipage}\hfill \\
               
               \begin{minipage}[b]{0.25\hsize}
                 \centering
                \includegraphics[keepaspectratio, scale=3]{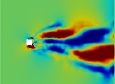}\hfill
                \subfloat{(b.1)~St~=~0.0456}
               \end{minipage} \hfill
               \begin{minipage}[b]{0.25\hsize}
                 \centering
                \includegraphics[keepaspectratio, scale=3]{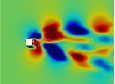}\hfill
                \subfloat{(b.2)~St~=~0.0923}
               \end{minipage} \hfill
               \begin{minipage}[b]{0.25\hsize}
                 \centering
                \includegraphics[keepaspectratio, scale=3]{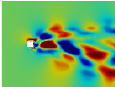}\hfill
                \subfloat{(b.3)~St~=~0.1134}
               \end{minipage} \hfill
               \begin{minipage}[b]{0.25\hsize}
                 \centering
                \includegraphics[keepaspectratio, scale=3]{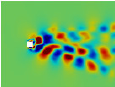}\hfill
                \subfloat{(b.4)~St~=~0.2061}
               \end{minipage} \hfill
             \end{tabular} \hfill
             \caption{Spatial distribution of DMD modes; a.1 $\sim$ a.4 are from conventional TDMD and b.1 $\sim$ b.4 are from Alternative TDMD}
             \label{ModeDist_cTDMDvsAltTDMD}
    \end{center}
\end{figure}

\subsection{Results of Incremental TDMD}
Next, Incremental TDMD is performed on the same flow field as the section 5.1 to investigate the relevancy of the algorithm comparing with the conventional TDMD algorithms.
However, as discussed in the section 5.1, Alternative TDMD resulted in the same DMD amplitudes and mode distributions as conventional TDMD.
Therefore, in this section, the result of Incremental TDMD is compared with the Alternative TDMD.
The difference between Alternative TDMD and Incremental TDMD is only input matrix which is computed by conventional SVD or incremental SVD, and all of the other algorithm is the completely same.
However, the Incremental TDMD can be performed on-the-fly, as singular values and vectors used for the algorithm can be computed on-the-fly, which is the big advantage with respect to the save of memory consumption and the usage of data storage, compared to Alternative TDMD.
However, in this paper, the incremental SVD is not performed in parallel to CFD simulation, but performed by using snapshot datasets saved in the storage, for the comparison with other TDMD algorithms.
The prescribed maximum rank of singular vectors and values for incremental SVD, thus for Incremental TDMD, is set to $r_{max} = 107$ which is the same rank as used in the computation in the section 5.1.
Resultantly, peak memory consumption for Incremental TDMD is 280MB compared to 1365MB for the Alternative TDMD computation, where Incremental TDMD saves almost 80\% memory for this case setups.
Please note that the Alternative TDMD program used for this computation is also very optimized with respect to the memory consumption, hence, it is expected that the regular implementation of the conventional TDMD algorithm seems to require much more memory.

First, the DMD amplitudes computed by the scaling with the first snapshot of flow field reconstructed from POD modes (explained in the section 2.3) are shown in Fig.5.5. with the Strouhal number of each mode. 
Comparing Fig5.5.(a) and (b), however, it seems to be difficult to find the relevancy of Incremental TDMD with respect to DMD amplitude distribution compared to the Incremental TDMD, even though it seems that some large amplitudes by Incremental TDMD at the similar frequency around~$St = 0.05,~0.1~\text{and}~0.2$ are also observed in amplitudes by Alternative TDMD.
In addition, it seems to be also difficult to identify the dominant modes by Incremental TDMD with respect to the magnitude of amplitude from Fig.5.5. 
%
%

\begin{figure}[htbp]
     \begin{tabular}{cc}
       \begin{minipage}[b]{0.5\hsize}
         \centering
        \includegraphics[keepaspectratio, scale=0.34]{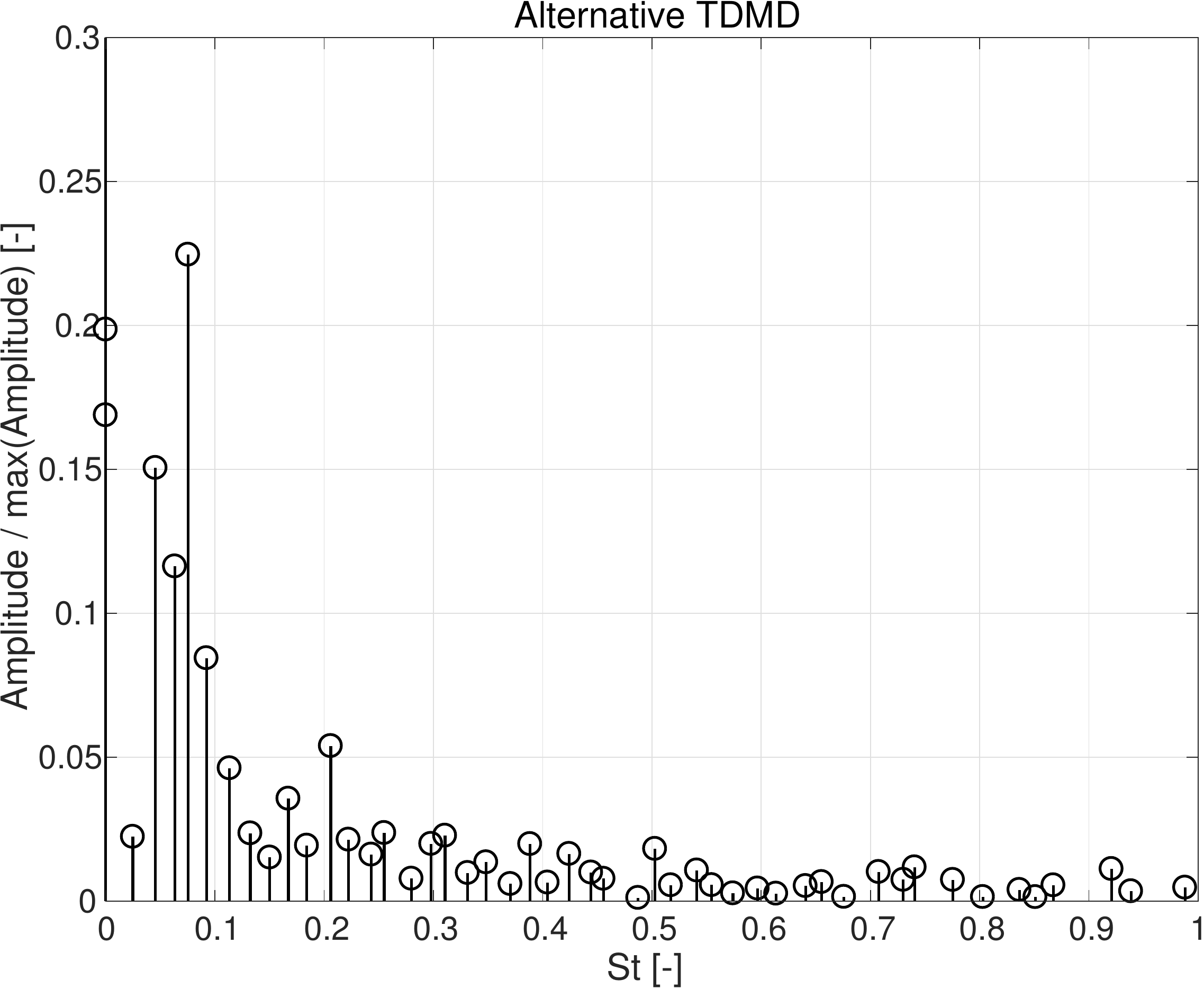}
        \subfloat{(a)~Alternative~TDMD}
       \end{minipage} &
       \begin{minipage}[b]{0.5\hsize}
         \centering
        \includegraphics[keepaspectratio, scale=0.34]{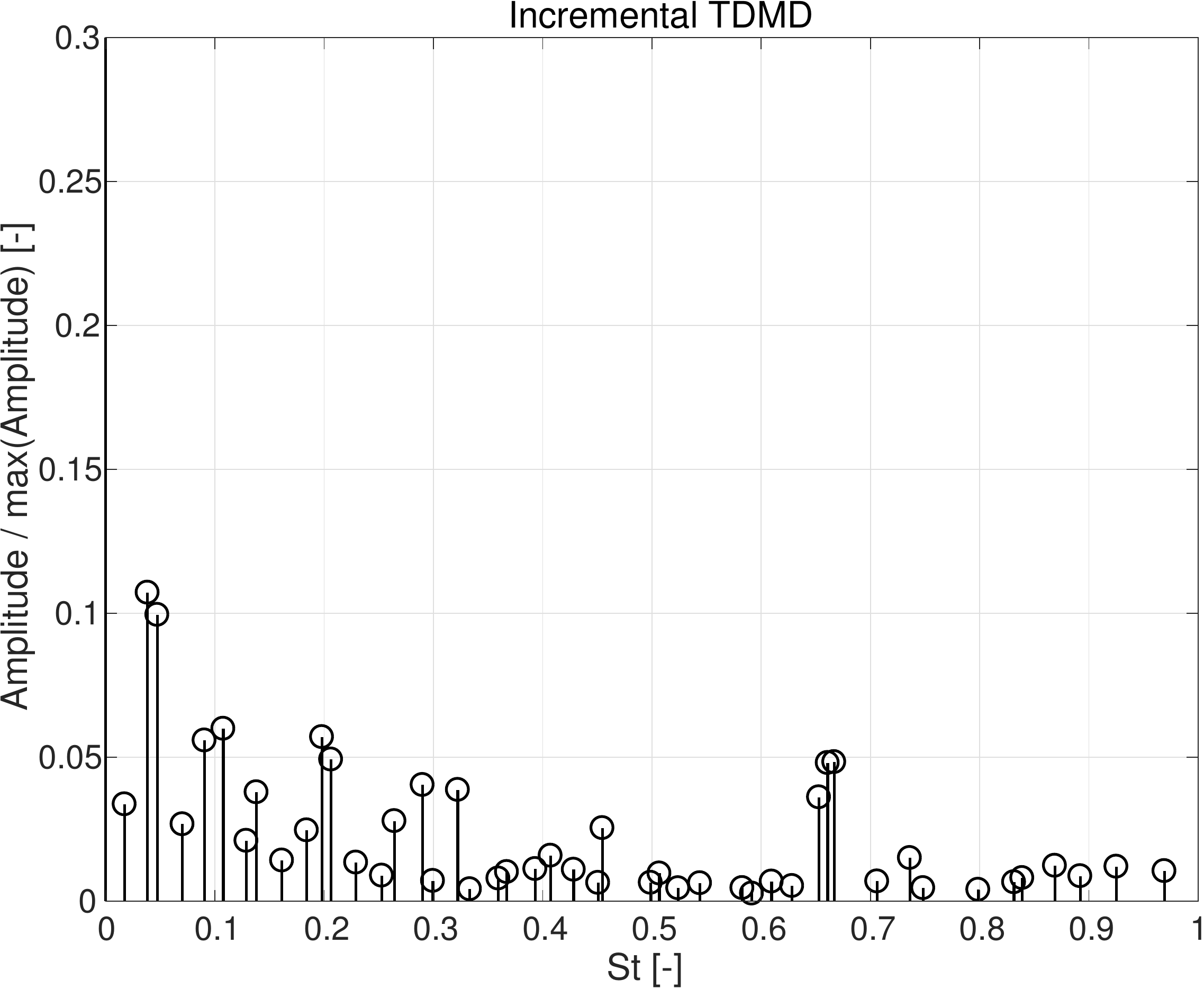}
        \subfloat{(b)~Incremental~TDMD}
       \end{minipage}
     \end{tabular}
         \caption{DMD~amplitude~by~the~first~snapshot~vector}
         \label{DMDmodeAmp_AltTDMDvsITDMD}
\end{figure}

Therefore, SPDMD is performed after Incremental TDMD and compared with the result of SPDMD with Alternative TDMD.  The SPDMD amplitudes are shown in Fig.5.6. with the optimal amplitude (represented as  Optimal amplitude) which is computed by Eq.~\eqref{eq:OptimalAmplitude} without the penalty parameter~$\gamma$.
In Fig.5.6., SPDMD amplitudes having different numbers of non-zero amplitudes are shown, which are induced by the different penalty parameter~$\gamma$ in Eq.(~\eqref{eq:MinNorm_SPDMDSP}.
Please note again that the numbers of modes having non-zero amplitudes are counted including negative frequency modes.
According to Fig.5.6, it is observed that the most of amplitude of SPDMD after Alternative TDMD at higher frequency than $St = 0.3$ become zero by SPDMD process, which is in agreement with SPDMD after Alternative TDMD.
However, the order of SPDMD mode by the magnitude of amplitude of each mode is the same regarding the frequency of mode, between Alternative TDMD and Incremental TDMD, even if remaining few modes by the SPDMD process. 
Nevertheless, the frequencies of remaining modes after SPDMD seem to be similar between Alternative TDMD and Incremental TDMD. 
%
%

\begin{figure}[htbp]
     \begin{tabular}{cc}
       \begin{minipage}[b]{0.5\hsize}
         \centering
        \includegraphics[keepaspectratio, scale=0.34]{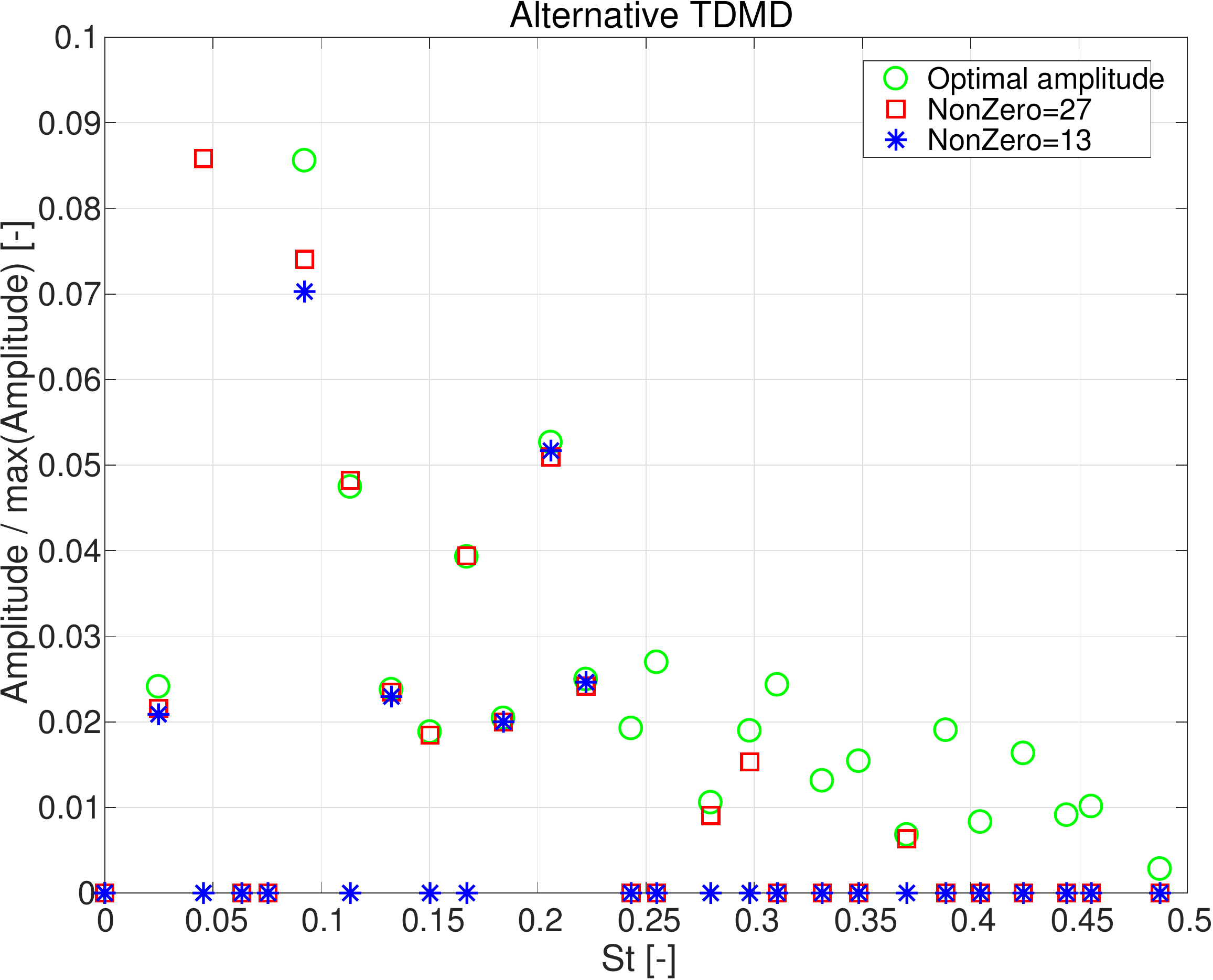}
        \subfloat{(a)~Alternative~TDMD}
       \end{minipage} &
       \begin{minipage}[b]{0.5\hsize}
         \centering
        \includegraphics[keepaspectratio, scale=0.34]{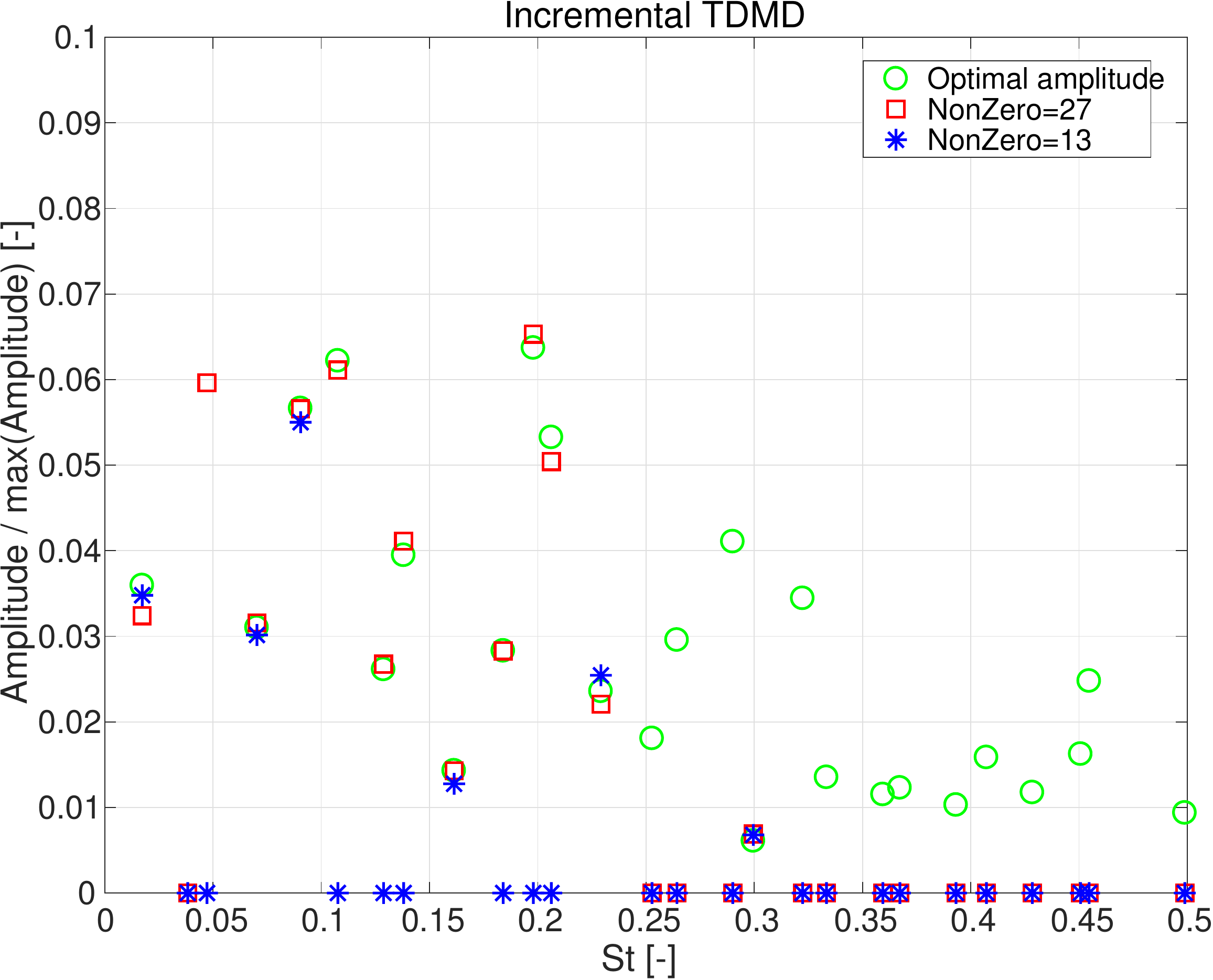}
        \subfloat{(b)~Incremental~TDMD}
       \end{minipage}
     \end{tabular}
         \caption{SPDMD~amplitudes~and~the~optimal~amplitude}
         \label{SPDMDmodeAmp_AltTDMDvsITDMD}
\end{figure}

For more detailed discussion, the SPDMD amplitudes when the number of non-zero amplitudes is 27 are shown in Fig.5.7, which are same as the amplitude distribution described as red square marker in Fig.5.6.
As shown in Fig.5.7.(a), the SPDMD amplitude after Alternative TDMD are comparably large at $St =$ 0.0456, 0.0923, 0.1134 and 0.2061 compared to the other amplitudes.
In contrast, comparably large amplitudes of SPDMD after Incremental TDMD are observed at $St = $ 0.0472, 0.0905, 0.1077 and 0.2063 as shown in Fig.5.7.(b), which are similar to SPDMD after Alternative TDMD with respect to the frequency, even though the order of amplitude by the magnitude is different, and another large amplitude is also observed at~$St = 0.1989$ which is not appeared in the result of Alternative TDMD.
Therefore, it seems that the similarity in SPDMD amplitude distribution between after alternative DMD and after Incremental TDMD is observed, with respect to the identification of the dominant DMD modes. 
%
%

\begin{figure}[htbp]
     \begin{tabular}{cc}
       \begin{minipage}[b]{0.5\hsize}
         \centering
        \includegraphics[keepaspectratio, scale=0.42]{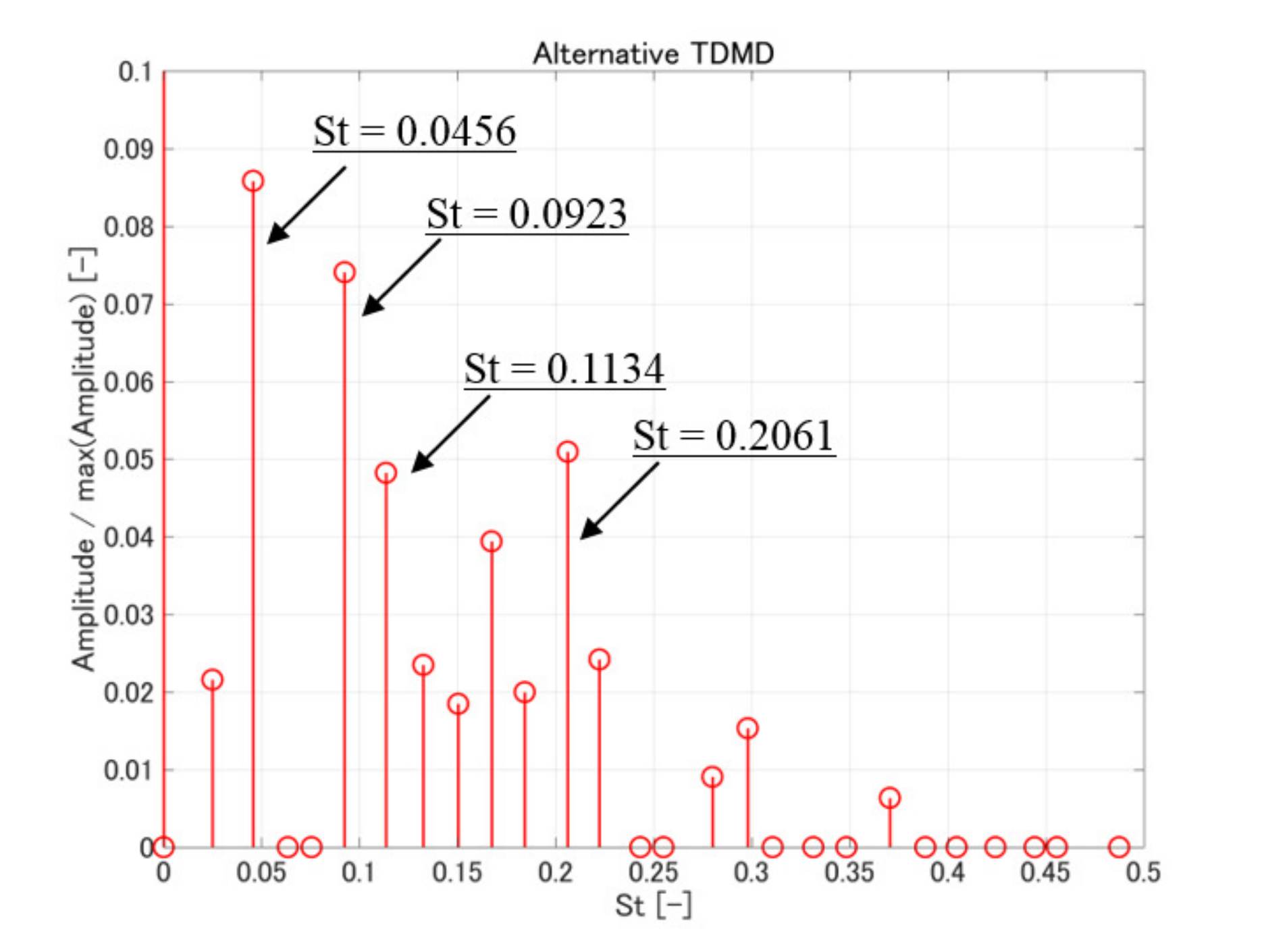}
        \subfloat{(a)~Alternative~TDMD}
       \end{minipage} &
       \begin{minipage}[b]{0.5\hsize}
         \centering
        \includegraphics[keepaspectratio, scale=0.42]{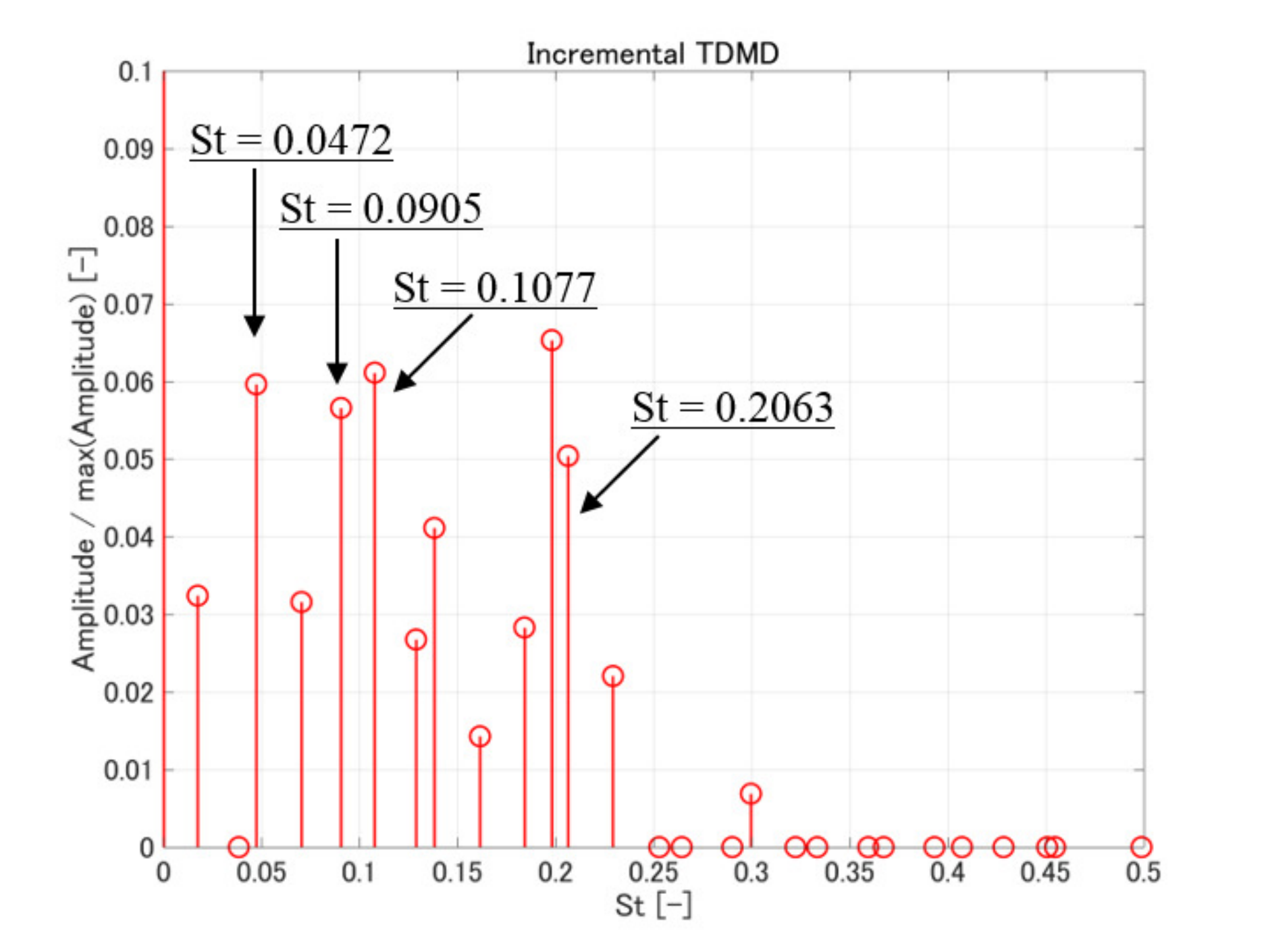}
        \subfloat{(b)~Incremental~TDMD}
       \end{minipage}
     \end{tabular}
         \caption{SPDMD amplitude (27 modes having non-zero amplitude)}
         \label{SPDMDmodeAmp_AltTDMDvsITDMD_27modes}
\end{figure}

The spatial distributions of DMD modes are visualized where the similarity is observed between Alternative TDMD and Incremental TDMD with respect to the mode amplitude and frequency of mode (Fig.5.8).
Comparing Fig.5.8.(a.2) and (b.2), it seems that those mode describes the same flow structure from the flow field at~$St \approx 0.09$.
Additionally, the mode distribution at~$St \approx 0.11$ shown in Fig.5.8.(a.3) and (b.3) also seem to be similar as well.
Therefore, the same flow structure from unsteady flow field seem to be identified by SPDMD combined with either Alternative TDMD or Incremental TDMD.
However, comparing Fig.5.8.(a.1) and (b.1), mode distribution~$St \approx 0.046$ seems to be similar, but the discrepancy is observed behind the square cylinder. Mode distribution at~$St \approx 0.206$ also shows slight discrepancy, comparing Fig.5.8.(a.4) and (b.4).

Though, the discrepancies appeared at~$St \approx 0.046$ and~$St \approx 0.206$ seem to be explained by the decomposition of a certain flow structure to a pair of mode.
According to the Fig.5.7.(b), for example, the high amplitude is observed at~$St = 0.1989$ neighboring to the mode at~$St = 0.2063$, which is not shown in the result of Alternative TDMD.
Therefore, mode distribution at~$St = 0.1989$ and~$St = 0.2063$ by Incremental TDMD are shown in Fig.5.9 and compared with the mode from Alternative TDMD.
It seems that Incremental TDMD mode at $St = 0.1989$ (Fig.5.9.(c)) and at~$St = 0.2063$  (Fig.5.9.(b)) are similar, and those mode seem to describe the same flow structure which is expressed by Alternative TDMD at~$St = 0.2061$ (Fig.5.9.(a)).
Accordingly, it seems that the slight difference at~$St \approx 0.046$  and~$St \approx 0.206$ caused the decomposition of a mode by Alternative TDMD to a pair of modes by Incremental TDMD, which seems to be induced by the condition of input matrix for the DMD operator computation.
In addition, if a certain, one mode is decomposed and expressed by a pair of modes in the other DMD computation, each mode of a pair can have smaller contribution to the total reconstructed flow field than expressed by the one mode, which may result in the different amplitudes by SPDMD where the optimization is conducted based on the contribution of each mode to the total energy of the flow field.
%
%

%
\begin{figure}[htbp]
    \begin{center}
             \begin{tabular}{cccc}
               \begin{minipage}[b]{0.25\hsize}
                 \centering
                \includegraphics[keepaspectratio, scale=3]{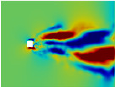}\hfill
                \subfloat{(a.1)~St~=~0.0456}
               \end{minipage} \hfill
               \begin{minipage}[b]{0.25\hsize}
                 \centering
                \includegraphics[keepaspectratio, scale=3]{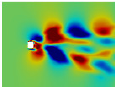}\hfill
                \subfloat{(a.2)~St~=~0.0923}
               \end{minipage} \hfill
               \begin{minipage}[b]{0.25\hsize}
                 \centering
                \includegraphics[keepaspectratio, scale=3]{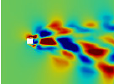}\hfill
                \subfloat{(a.3)~St~=~0.1134}
               \end{minipage} \hfill
               \begin{minipage}[b]{0.25\hsize}
                 \centering
                \includegraphics[keepaspectratio, scale=3]{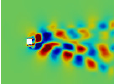}\hfill
                \subfloat{(a.4)~St~=~0.2061}
               \end{minipage}\hfill \\
               
               \begin{minipage}[b]{0.25\hsize}
                 \centering
                \includegraphics[keepaspectratio, scale=3]{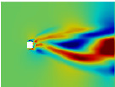}\hfill
                \subfloat{(b.1)~St~=~0.0472}
               \end{minipage} \hfill
               \begin{minipage}[b]{0.25\hsize}
                 \centering
                \includegraphics[keepaspectratio, scale=3]{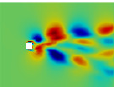}\hfill
                \subfloat{(b.2)~St~=~0.0905}
               \end{minipage} \hfill
               \begin{minipage}[b]{0.25\hsize}
                 \centering
                \includegraphics[keepaspectratio, scale=3]{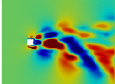}\hfill
                \subfloat{(b.3)~St~=~0.1077}
               \end{minipage} \hfill
               \begin{minipage}[b]{0.25\hsize}
                 \centering
                \includegraphics[keepaspectratio, scale=3]{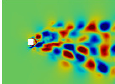}\hfill
                \subfloat{(b.4)~St~=~0.2063}
               \end{minipage} \hfill
             \end{tabular} 
             \caption{Spatial distribution of DMD modes; a.1 $\sim$ a.4 are from Alternative TDMD and b.1 $\sim$ b.4 are from Incremental TDMD}
             \label{ModeDist_altTDMDvsITDMD}
    \end{center}
\end{figure}
%
%

\begin{figure}[htbp]
     \begin{tabular}{ccc}
       \begin{minipage}[b]{0.33\hsize}
         \centering
        \includegraphics[keepaspectratio, scale=4.5]{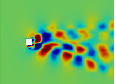}\hfill
        \subfloat{(a)~St~=~0.2061}\\
        \subfloat{(Alternative~TDMD)}
       \end{minipage} \hfill
       \begin{minipage}[b]{0.33\hsize}
         \centering
        \includegraphics[keepaspectratio, scale=4.5]{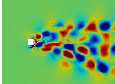}\hfill
        \subfloat{(b)~St~=~0.2063}\\
        \subfloat{(Incremental~TDMD)}
       \end{minipage} \hfill
       \begin{minipage}[b]{0.33\hsize}
         \centering
        \includegraphics[keepaspectratio, scale=4.5]{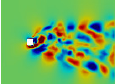}\hfill
        \subfloat{(c)~St~=~0.1989}\\
        \subfloat{((Incremental~TDMD)}
       \end{minipage} \hfill
     \end{tabular}
         \caption{Spatial distribution of DMD modes at St~$\approx$~0.2}
         \label{ModeDist_altTDMDvsITDMD_St0pt2}
\end{figure}

In conclusion, it is revealed that the combination of SPDMD and Incremental TDMD computes relevant modes compared to SPDMD and Alternative TDMD, thus also relevant to SPDMD and conventional TDMD. 
Discrepancies in SPDMD amplitude distribution and spatial distribution of mode can be explained.
Therefore, it seems that the combination of SPDMD and Incremental TDMD is applicable to the identification of the dominant flow structure from large datasets, such as used in engineering applications.
%
%

\section{Conclusion}
In this paper, first, an alternative algorithm of TDMD (Alternative TDMD) that is computed by only singular values and vectors of snapshot matrix is proposed, where Singular Value Decomposition (SVD) in the algorithm is performed by conventional SVD.
For validation purpose, Alternative TDMD is applied on the velocity magnitude around the infinite square cylinder, which results in the same DMD amplitudes compared to conventional TDMD.
Additionally, SPDMD is conducted after the Alternative TDMD, which also results in the amplitude as SPDMD after conventional TDMD.
Spatial distributions of dominant DMD modes identified by SPDMD are also same between conventional TDMD and Incremental TDMD.
Therefore, it is validated that Alternative TDMD combined with SPDMD results in totally same DMD amplitudes and mode distributions as conventional TDMD with SPDMD.

Second, the Alternative TDMD algorithm is performed with incremental SVD instead of conventional SVD (Incremental TDMD).
The result of Incremental TDMD is compared with the result of Alternative TDMD.
Additionally, SPDMD is also performed after Incremental TDMD and Alternative TDMD.
Comparing the DMD mode amplitude computed by scaling to the first snapshot vector, it is difficult to find the relations between Incremental TDMD and Alternative TDMD.
However, some dominant modes by Incremental TDMD can be found in the similar frequency as Alternative TDMD, by comparing the result of SPDMD, even though some discrepancies are still found in SPDMD mode amplitude distribution.
Therefore, it is revealed that SPDMD enable to find the dominant modes computed by Incremental TDMD which is relevant to the modes by Alternative TDMD.
The spatial distribution of dominant modes by Incremental TDMD are visualized and compared with those by Alternative TDMD, which result in the similar spatial mode distribution at the close frequencies even though there are some discrepancy.
However, it seems that the discrepancies can be explained as a certain mode is decomposed to a pair of modes, which result in the difference in amplitude and spatial mode distribution.

In conclusion, Incremental TDMD combined with SPDMD seems to result in relevant SPDMD amplitude and spatial mode distribution compared to the Alternative TDMD and conventional TDMD combined with SPDMD. 
Therefore, it seems that Incremental TDMD with SPDMD is useful for the analysis of huge and complex dataset of the flow field, with respect to the identification of the dominant flow structure by using the small computational resources, as Incremental TDMD doesn’t require saving snapshot data on the storage and loading all dataset to the memory space.

\section*{Appendix.A}
In the practical computation, performing SVD requires huge computational resources with respect to memory consumption and computational time. Therefore, singular vectors and values of snapshot matrices are computed by using the method of snapshot~\cite{Sirovich1987} in this paper, instead of performing conventional SVD. The algorithm of conventional DMD performed with the method of snapshot is described in the followings, as an example.

Instead of performing SVD in Eq.\eqref{eq:SVD_X}, the eigendecomposition of auto-covariance matrix of snapshot matrix~${\bm X}$ is firstly performed as follows.
%
\begin{equation}
  ({\bm{X}}^{T} {\bm{X}}) {\bm{W}} = {\bm{W}} {\bm{\Sigma}}^{2}
  \tag{A.1}
  \label{eq:EigDecomposeXtX}
\end{equation}
where eigenvectors become the right singular vectors of~${\bm X}$ , and eigenvalues become square of singular values of~${\bm X}$. Additionally, left singular vector of~${\bm X}$, which is same as POD modes of~${\bm X}$ by the method of snapshot, is computed from right singular vectors and singular values, as follows.
\begin{equation}
  {\bm U} = {\bm X} {\bm W} {\bm \Sigma}^{-1}
  \tag{A.2}
  \label{eq:RightSingularVectorX}
\end{equation}

Hence, the projected DMD operator~${\bm{\tilde{A}}}$ in Eq.\eqref{eq:PrjDMDOperator_Schmid} is computed as follows.
\begin{equation}
  {\bm{\tilde{A}}} = {\bm U}^{T} {\bm A} {\bm U} = {\bm \Sigma}^{-1} {\bm W}^{T} {\bm X}^{T} {\bm Y} {\bm W} {\bm \Sigma}^{-1}
  \tag{A.3}
  \label{eq:DefPrjDMDOperator_Snapshot}
\end{equation}

Finally, DMD modes are computed can be computed as follows.
\begin{equation}
  {\bm \hat{\Phi}} = {\bm U} {\bm V} = {\bm X} {\bm W} {\bm \Sigma}^{-1} {\bm V}
  \tag{A.4}
  \label{eq:unscaledDMDmode_Snapshot}
\end{equation}
where eigendecomposition of projected DMD operator~${\bm{\tilde{A}}}$ is computed as follows.
\begin{equation}
  {\bm{\tilde{A}}} {\bm {V}} = {\bm {V}} {\bm \Lambda}
  \tag{A.5}
  \label{eq:eig_prjAtilde_appendix}
\end{equation}

In this algorithm, it is not necessary to compute the left singular vectors~${\bm U}$ explicitly, which results in the more efficient computation with respect to the memory consumption and computational time. 

In addition, the scaling factor of DMD modes explained in the section 2.3 can be computed more efficiently. Firstly, the first snapshot vector~${\bm x}_0$ is reconstructed from the first column of the transposed right singular vectors~${\bm W}^{T} = \{{{\bm w}_{0}}^{T}, {{\bm w}_{1}}^{T}, \ldots, {{\bm w}_{m-1}}^{T}\}$, as follows.
\begin{equation}
  {\bm x}_{0} = {\bm U} {\bm \Sigma} {\bm w}_0 ^{T}
  \tag{A.6}
  \label{eq:firstSnapshot_Recons}
\end{equation}
Therefore, the Eq.\eqref{eq:Solution_LinearEq_Scaling} can be modified as shown in the following.
\begin{equation}
\begin{split}
  {\bm d} &= {\bm {\hat{\Phi}}}^{+} {\bm x}_{0}\\
    &= ( {{\bm V}^{-1}} {\bm \Sigma} {\bm W}^{T} {\bm X}^{+}) \left( ( {\bm X} {\bm W} {\bm{\Sigma}}^{-1} ) {\bm \Sigma} ~{{\bm w}_{0}}^{T} \right) \\
    &= {\bm V}^{-1} {\bm \Sigma} ~{{\bm w}_{0}}^{T}
\end{split}
    \tag{A.7}
    \label{eq:LinearEq_ScalSnap}
\end{equation}

In the equation~\eqref{eq:LinearEq_ScalSnap}, the DMD amplitude is more efficiently computed with respect to the usage of computational resources, as the pseudo-inverse of huge matrix~${\bm{\hat{\Phi}}}$ is not necessary to be solved.


\end{document}